\documentclass[10pt,letterpaper]{article}
\usepackage[top=0.85in,left=2.75in,footskip=0.75in]{geometry}

\usepackage{changepage}

\usepackage[utf8x]{inputenc}

\usepackage{textcomp,marvosym}

\usepackage{amsmath,amssymb,bm}

\usepackage{cite}

\usepackage{nameref,hyperref}

\usepackage[right]{lineno}

\usepackage{microtype}
\DisableLigatures[f]{encoding = *, family = * }

\usepackage{color}
\usepackage{siunitx}
 
\raggedright
\usepackage[aboveskip=1pt,labelfont=bf,labelsep=period,justification=raggedright,singlelinecheck=off]{caption}

\makeatletter
\renewcommand{\@biblabel}[1]{\quad#1.}
\makeatother

\date{}

\usepackage{lastpage,fancyhdr,graphicx}
\usepackage{epstopdf}
\fancyheadoffset[L]{2.25in}
\fancyfootoffset[L]{2.25in}

\usepackage{multirow}

\begin{document}
\vspace*{0.2in}

\begin{flushleft}
{\Large
\textbf\newline{Collective Decision Dynamics in Group Evacuation: Behavioral Experiment and Machine Learning Models} 
}
\newline
\\
Chantal Nguyen\textsuperscript{1}*,
Fangqiu Han\textsuperscript2,
Kimberly J. Schlesinger\textsuperscript{1},
Izzeddin G{\"u}r\textsuperscript{2},
Jean M. Carlson\textsuperscript{1}
\\
\bigskip
\textbf{1} Department of Physics, University of California, Santa Barbara, Santa Barbara, California, United States of America
\\
\textbf{2} Department of Computer Science, University of California, Santa Barbara, Santa Barbara, California, United States of America
\\
\bigskip

* cnguyen@physics.ucsb.edu

\end{flushleft}
\frenchspacing
\section*{Abstract}
Identifying factors that affect human decision making and quantifying their influence remain essential and challenging tasks for the design and implementation of social and technological communication systems. We report results of a behavioral experiment involving decision making in the face of an impending natural disaster. In a controlled laboratory setting, we characterize individual and group evacuation decision making influenced by several key factors, including the likelihood of the disaster, available shelter capacity, group size, and group decision protocol. Our results show that success in individual decision making is not a strong predictor of group performance. We use an artificial neural network trained on the collective behavior of subjects to predict individual and group outcomes. Overall model accuracy increases with the inclusion of a subject-specific performance parameter based on laboratory trials that captures individual differences. In parallel, we demonstrate that the social media activity of individual subjects, specifically their Facebook use, can be used to generate an alternative individual personality profile that leads to comparable model accuracy. Quantitative characterization and prediction of collective decision making is crucial for the development of effective policies to guide the action of populations in the face of threat or uncertainty.

\section*{Introduction}

Collective decisions among small groups of humans in a crisis are difficult to predict because of the inherent complexity of human behaviors and interactions. Disasters and other threat situations often display noisy and fragile individual performance that can lead to non-optimal collective outcomes \cite{stir7} \cite{stir8} \cite{evachouse} \cite{stir10}. Groups can exhibit a tendency to make riskier collective decisions after group discussion than individuals would alone \cite{stoner} \cite{isenberg}, and social interactions can lead to a ``mob mentality'' \cite{stir12} \cite{stir13} \cite{stir14} that hinders evacuation and may result in injury and violence. Associated spatiotemporal clustering of departure times can lead to traffic congestion and delays \cite{stir15} \cite{stir16} \cite{stir17}. Quantifying synergies between broadcast information involving the objective status of a threat and concurrent social factors that may inhibit or accelerate action is critical to devising effective strategies for ensuring the safety of populations in crisis situations. 

Challenges to coordinating collective action when a population is at risk may be exacerbated by a variety of influences, including the desire to locate a missing family member or panic of an individual group member, each of which may derail action for the entire group \cite{katrina1} \cite{katrina2}. Alternatively, desirable outcomes occur when group action exceeds the sum of its parts. Identifying and promoting effective teams has been investigated in the context of emergent leadership, wisdom of crowds, performance stability, and division of labor \cite{munger} \cite{emerglead} \cite{decisionsurvey}. Central questions include the following: to what extent does individual behavior predict that individual's performance in a group? How does the size, diversity, communication, and decision mechanism of a group impact their collective performance? To what extent can individual performance differences within a group be deduced from other (social network) measurable behaviors \cite{munger} \cite{groupreview}?

This paper isolates and quantifies tensions in decision making that arise when individuals act cohesively, as in families, teams, or squads. We compare the behavior of individuals acting alone with their behavior in groups, whose actions are linked through a network of influence or communication in a crisis situation. In this paper, we employ a combination of experiments in behavioral network science, analysis of social network usage, and artificial neural network models to quantify the consequences of different protocols for decision making among groups that are part of a larger population subject to an impinging crisis. 
Our results link features of individual human behavior with observable behaviors of groups and thereby address critical issues that arise in identifying individual differences that constrain or promote achievement of group objectives \cite{munger} \cite{groupreview}. The insights gained from investigating collective behavior in evacuation scenarios through behavioral experiments and data-driven modeling can inform the design of public policies for collective action in stressful and uncertain situations.

\textbf{Our work in context.} Our experimental approach of quantifying specific behavioral drivers in a controlled setting constitutes an intermediate between surveys conducted in the aftermath of disasters, which can be difficult to execute on a large scale and often qualitative, and controlled human-subject experiments of isolated cognitive tasks.
In this controlled setting, all information observed by participants prior to their decisions is recorded, facilitating data-driven modeling of decision making.
Our approach capitalizes on prior work developing mathematical models of information flow on social networks to isolate drivers of collective dynamics \cite{dani} and performing behavioral experiments to quantify individual decision making in crisis situations \cite{sean}. The focus of these previous studies has been on evacuate-or-not responses to simulated natural disasters studied through a model-experiment-model paradigm, where initial theoretical studies predict effects of decision making under various social and economic pressures, followed by experimental tests of such predictions to quantify those effects. The analysis of such experiments enables data-driven models which in turn inform the design of subsequent experiments. 

A theoretical component of this previous work \cite{sean} identified a decision model for evacuation behavior that described the heterogeneous and non-optimal behavior of human subjects observed empirically in behavioral experiments which formed the basis for the experiment described in this paper. The model was used to quantify individual differences in decision dynamics and relate those differences to psychometric measurements of personality and risk attitudes. This model was derived from detailed empirical observations, and as such stands in sharp contrast to a set of models typically used in numerical simulations or large-scale, data-driven studies that treat decisions as random, optimal, or based on a threshold applied to a state variable representing opinion, which is updated by an assumed interaction rule (e.g., \cite{stir12} \cite{stir13} \cite{stir17} \cite{dani} \cite{stir18} \cite{stir19}).

In our current study, we probe mesoscale decision and behavior dynamics by dividing members of a community into cohesive groups. Each group must act as a unit and make a forced evacuate-or-not decision in the face of a disaster, which will either hit or miss the community according to a time-evolving likelihood. Building on the earlier work involving tradeoffs between broadcast and social information on individual decision making \cite{sean}, this project focuses specifically on isolating novel features that arise when decisions impact a group. 

Our experimental platform adapts the framework of Kearns et al. \cite{kearns1} \cite{kearns2} \cite{stir5} \cite{stir6}, who have conducted a series of ``behavioral network science'' (BNS) experiments that have focused on collective problem solving tasks, such as abstract graph coloring problems or economic investment games. In both competitive and cooperative tasks and scenarios, human subjects in these previous BNS studies have been shown to ``perform remarkably well at the collective level'' \cite{stir6}. In contrast to the work of Kearns et al., our investigations involve decision making in a threat scenario where we expect both heterogeneous behaviors across individual subjects and non-optimality of behavior at both the individual and collective levels.

\textbf{Experimental overview.}
The experiment described in this paper places participants in a sequence of simulated disaster scenarios, each of which involves an approaching threat (e.g., wildfire or hurricane) that may impact their community within an uncertain window of time. The status of the threat is described by a time-dependent likelihood based on an underlying stochastic process. Resource limitations are represented by a finite number of available spaces in an evacuation shelter. Participants are at risk of losing a portion of their initial wealth based on their action and the disaster outcome. In each scenario, individuals make a binding evacuate-or-not decision based on the information available. 

In some scenarios, participants act as individuals, in which case, once the decision to evacuate is made, the individual action (evacuation to the shelter) follows immediately, providing sufficient shelter space is available. In other scenarios, the same individuals are grouped into teams. In these cases, group action is determined by the collective decisions of the team subject to availability of shelter space and a forced action protocol. The action protocols used in this study include majority-vote (the group evacuates when the majority has decided), first-to-go (the group evacuates when the first member decides), and last-to-go (the group evacuates when the entire group achieves consensus). This experimental design enables us to quantitatively compare and contrast individual and group decision making, as well as the effectiveness of different protocols for combining individual decisions into collective group action. 

For example, under the first-to-go protocol, in which a ``go'' decision made by one individual within a group forces the entire group to evacuate, behaviors observed in individual humans without a group influence (such as tending to evacuate earlier than is optimal \cite{sean}) would result in poor population-level performance: a single individual's early evacuation commits the entire group to action, resulting in collective loss of temporal and financial resources. For optimal group performance in a first-to-go scenario, individuals must instead modify their behavior, either by reducing evacuation rates, increasing disaster likelihood thresholds for evacuating, or by deferring to other members of the group to make the decisions. 

Finally, based on the results of the behavioral experiment, we develop individual data-driven artificial neural network models for decision making which incorporate individual and group disaster scenarios. Heterogeneity in the decision models is linked to individual differences, identified using two alternate means. The first is based directly on the experimental data. The second is determined by social network (specifically Facebook) use. The two methods produce quantitatively similar decision models and overall accuracy, suggesting that social network use may be predictive of individual behavior in a group setting. 

Differences in personality traits and risk perception can influence decision making, specifically in scenarios involving evacuation from natural disasters \cite{evachouse}. Common assessments of personality include the Big Five Inventory questionnaire, a questionnaire based on five characteristics (extraversion, agreeableness, conscientiousness, neuroticism, and openness) often used in psychological research \cite{bigfivedigman}\cite{bigfivemccrae}, and common assessments of risk perception include the Domain-Specific Risk-Attitude Scale \cite{risk1}\cite{risk2}. Furthermore, the rapid development and widespread adoption of social networks such as Facebook and Twitter has lead to increasing interest in identifying personality-related factors from online profiles and social media activity \cite{fb}. Previous studies have commonly focused on developing predictive personality profiles from individual behavior on social media platforms, and it has been shown that computer models can outperform humans in assessing personality \cite{fb} \cite{fblikes}.

The experiment detailed in this paper facilitates a rich investigation of the differences between group and individual behaviors and the sub-optimality of human behavior. A further exploration of collective decision dynamics is described in \cite{paper2}, which expands the analysis of the empirical data presented here and uses a complementary but methodologically different approach to modeling decision making. 
In \cite{paper2}, a few-parameter decision model with an explicit functional form motivated by empirical data is used to model individual decisions, simulate group behavior, and determine optimal decision strategies with Bayesian inference.

In this paper, our inherently multiscale results link characteristics of individual humans, observed behaviors of groups, and outcomes for the population as a whole, thereby addressing issues that arise in identifying individual differences that constrain or promote group objectives. We obtain information about the role of cohesive units in decision making, which is correlated with individual differences in the population, and which identifies emergent leaders as individuals willing to make decisions for the group, as well as followers who avoid initiating group action. The results have direct relevance to complex, real-world threat scenarios as well as coordination and control of civilian populations at risk.

\section*{Methods}
\subsection*{Evacuation Experiment}
\begin{figure}[ht!]
\centering
\includegraphics[scale=1]{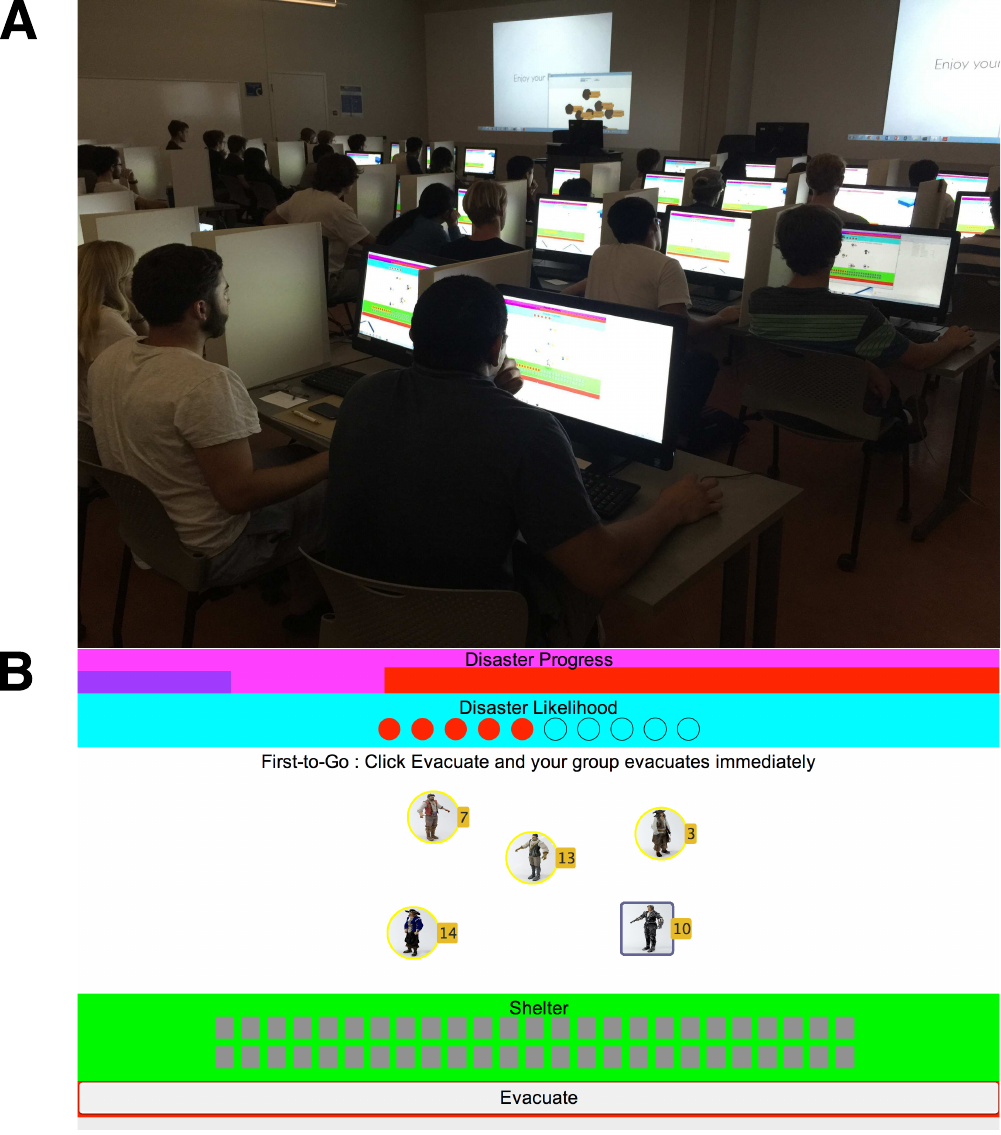}
\caption{\textbf{An overview of the evacuation experiment.} {A: Experimental setup. 50 participants were provided with information on an impending virtual natural disaster via an individualized computer interface.  Dividers were placed between each computer.}
B: Sample computer interface. This example is a first-to-go trial with 5-person groups. The Disaster Progress bar at the top of the page shows the time progress of the trial. The Disaster Likelihood bar displays the time-dependent probability of the disaster strike, represented by the proportion of filled red circles; {five} filled circles indicates a likelihood of {5} out of 10.
The white center panel indicates the evacuation protocol for group trials, first-to-go, as well as a visualization of the participant's group members, represented by their distinctive avatars. The number to the right of each avatar indicates the corresponding participant's rank.
The green Shelter bar represents the evacuation shelter, which fills up as participants evacuate; available spaces are indicated by gray rectangles. Participants click the Evacuate button at the bottom of the screen when they decide to evacuate.}
\label{fig:overview}
\end{figure}

On March 13, 2015, we conducted a controlled behavioral experiment at the University of California, Santa Barbara (UCSB) in order to quantitatively characterize the decision-making behavior of both individuals and groups in stressful situations. In our experiment, 50 participants each decided if and when to evacuate from an impending virtual natural disaster (Fig \ref{fig:overview}A). All participants provided written informed consent, and the experimental protocol was approved by the Institutional Review Board of UCSB.

Our experimental design is based on that of Carlson et al. \cite{sean} but differs primarily with the addition of groups. In both experiments, 50 individuals participated in several scenarios (trials) where the strike likelihood of an impending virtual disaster was broadcast at regular intervals, along with the availability of beds in an evacuation shelter. Participants used this information to decide if and when to evacuate. They were staked a number of points before each trial, and points were deducted from the initial stake based on the success of their actions. While participants in the previous study made decisions as individuals, our experiment also includes scenarios where participants were randomly assigned to groups which act according to a specified protocol.

This experiment involved 144 trials, each lasting 20 to 60 half-second time steps. During each trial, participants were provided with \textit{disaster} information about the progression and likelihood of the disaster hitting the community and the occupation status of an evacuation shelter via an individualized computer interface, an example of which is shown in Fig \ref{fig:overview}B. Participants began each trial in the \textit{at home} state and, upon evacuating from the disaster, entered the \textit{in shelter} state.

At the beginning of each trial, participants were each staked 10 points. The number of points deducted from this stake was a function of whether or not a participant evacuated successfully and whether or not the disaster struck, as shown in the loss matrix in Table \ref{table:lossmat}. Participants were ranked after each trial by their cumulative score and were awarded monetary compensation at the end of the experiment based on their final score over all trials.

\begin{table}[h!] \centering
\caption{\textbf{Point Values at Stake for Each Trial.}}
\normalsize
\begin{tabular} {|l|r|r|}
\hline
\multicolumn{3}{|c|}{\textbf{Points deducted}}\\
\hline
 & Disaster hits & Disaster misses  \\ \hline
At home  & -10 & 0 \\ \hline
In shelter  & -6 & -2 \\
\hline
\multicolumn{3}{|c|}{\textbf{Net points gained}}\\
\hline
 & Disaster hits & Disaster misses  \\ \hline
At home  & 0 & 10 \\ \hline
In shelter  & 4 & 8 \\
\hline
\end{tabular}
\caption*{Participants are initially staked 10 points in each trial. They lose 6 or 10 points if the disaster hits and 0 or 2 points if the disaster misses, depending on whether they evacuated or remained home.}
\label{table:lossmat}
\end{table}

In this experiment, we focus on the effects of both broadcast information and \textit{social} information on decision making. In some trials, participants could only evacuate as individuals; in others, participants were randomly assigned to groups of five or 25 individuals each which collectively act according to one of three specified protocols: first-to-go (``FTG''), majority-vote (``MV''), and last-to-go (``LTG''). The ultimate action of the group was dependent on the combined decisions of individuals within the group. Participants could view the decisions and rankings of their group members, referred to as \textit{social} information, to formulate their strategy based on their group-mates' behavior.

Prior to the behavioral experiment, we collected each participant's Facebook archive, which contains their profile, liked pages, and site activity. We later used this information to formulate a measure of their personality. As shown in Table \ref{table:face}, we collected data on the variables of \textit{gender}, \textit{age}, \textit{number of friends}, and \textit{page likes} of films, television programs, and books. Given the diversity and quantity of Facebook pages, Facebook likes can constitute an individual's unique ``digital footprint'' and can also reflect other Internet behavior such as web searches or online purchases \cite{fb}. Furthermore, Facebook likes have been shown to correlate with personality assessments based upon the five-factor, or Big Five, model \cite{bigfivedigman} \cite{bigfivemccrae} \cite{fb} \cite{fblikes}.

\begin{table}
\caption{\textbf{List of Attributes Extracted from Facebook Data and the Ranges of Their Values.}}
\centering    
\normalsize
		\begin{tabular}{| c | c |}
			\hline
			\textbf{Data} & \textbf{Range} \\ \hline
			Gender &  \{Female, Male, NS/Other\}\\
            Age & 19 - 31\\ 
            \# of Facebook friends & 8 - 1956\\     \hline
            \textbf{Top genres from movie, TV} & \\ 
            \textbf{shows, and book pages} & \textbf{Range} \\ \hline
            War (novel) & 0 - 27 \\
            Young adult fiction & 0 - 19 \\
            Romance (novel) & 0 - 15 \\
            Realistic fiction & 0 - 14 \\
            Mystery & 0 - 6 \\  \hline 
           
		\end{tabular}
		
\caption*{We extract demographic information (gender -- female, male, not specified (NS) or other -- and age), number of friends, and ``liked'' pages from participants' Facebook archives, which are used to determine individualized personality profiles for each participant. Liked pages are grouped into genres; the top five genres are shown, along with the range of liked pages per genre per person.}		
    \label{table:face}
\end{table}

\begin{figure}[ht!]
\centering
\includegraphics[scale=1]{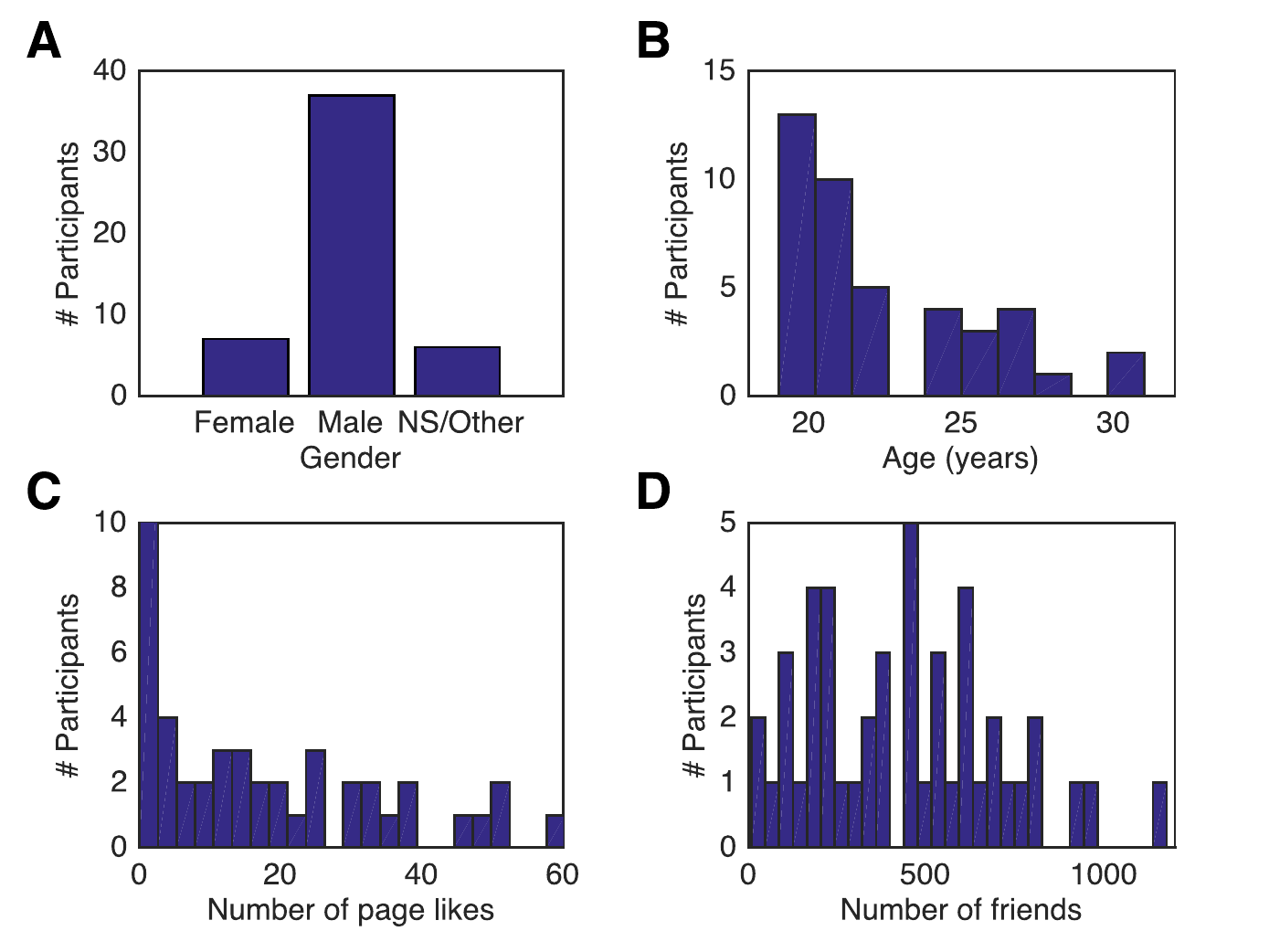}
\caption{\textbf{Statistics determined from participants' Facebook data.} A: Gender. 7 out of 50 participants identified as female while 37 out of 50 identified as male; the remaining 6 participants identified as another gender or did not specify their gender (NS/Other). B: Age distribution. The ages of most participants lie between 20 and 30, as most are university students. C: Distribution of number of Facebook page likes. Most participants have fewer than 50 total likes of Facebook pages. D: Distribution of number of Facebook friends.}
\label{fig:facestats}
\end{figure}

Statistics from the Facebook data are shown in Fig \ref{fig:facestats}.
Due to the wide range of liked pages and low overlap in specific page likes between individuals, we clustered movies, TV shows, and books according to genre, e.g., adventure, science fiction, and romance, for a total of 73 clusters. Movie and TV show genres were obtained from the Internet Movie Database (IMDb) \cite{imdb}, and book genres were obtained from Wikipedia \cite{wiki}. The cumulative number of likes in each cluster were used to determine personality metrics for each participant; these were then used to develop decision models which incorporate individual heterogeneity via one unique personality parameter per participant determined by training an artificial neural network on the data.

\subsection*{Nested Experimental Design}

To isolate key factors of decision making, we employ a principled method of exploring parameter variations which exploits statistical power. We use a nested experimental design which allows us to employ scenarios generated from all possible permutations of disaster likelihood, shelter capacity, group size, and group protocol, resulting in a wealth and diversity of parameter interactions. In our experimental setup, each user sits at an individual computer interface (Fig \ref{fig:overview}) in which {disaster} information and {social} information are displayed and updated at each time step. From this information, the participant decides whether or not to evacuate.

Our experimental design contains numerous deliberate choices regarding the presentation of information. Our choices are based on studies that report relationships between the method of displaying quantitative information and the resulting information extracted by humans. For example, the graphical presentation of probabilistic information, as well as whether the information is reported as discrete points or as a continuous distribution, has been shown to affect the accuracy of decisions based on this information \cite{edwards}. Since real-world information during a potential disaster scenario is never certain and often involves estimates and probabilities, we chose to display disaster information as a likelihood value at low resolution to reflect the limited amount of information available during disaster situations \cite{regnier}.

\textbf{Evolution of disaster likelihood.} The progression of the disaster is determined by an underlying stochastic process, previously described in \cite{phit}, in which a ``threat'' executes a generalized random walk towards a target. The disaster likelihood, denoted by $P_{hit}$, varies with time according to this stochastic process, where the threat moves in two dimensional space toward a target with random motion in the transverse direction and monotonic advancement in the longitudinal direction. The transverse and longitudinal motion both progress in stochastically varying step sizes within a specified range, such that the minimum time required for progressing from the origin to the longitudinal position of the target is 20 time steps, and the maximum time is 60 time steps. 
The disaster strikes (``Hit'') if the threat contacts the target, while the disaster misses (``Miss'') if the threat passes the longitudinal position of the target without hitting. The disaster likelihood $P_{hit}$ can take on continuous values between 0 and 1 and is computed at each time step. Once the threat reaches the longitudinal position of the target, the likelihood will either be exactly 0, indicating a miss, or 1, indicating a hit.

In a realistic natural disaster scenario, available information on the progression of the disaster is typically reported as a strike probability \cite{regnier}. In our experiment, participants are shown the $P_{hit}$ value as it updates, but the $P_{hit}$ value is rounded down to the nearest tenth before it is displayed, reflecting the limited information available during a realistic scenario. The full history of $P_{hit}$ values during each trial is not shown. At the beginning of each trial, $P_{hit}$ is initialized at a value between $0.3$ and $0.5$. When the actual $P_{hit}$ value becomes either 0 or 1, i.e., the threat reaches the longitudinal position of the target, the trial immediately ends. Since the $P_{hit}$ value displayed to participants has been rounded down to the nearest tenth, it is possible for participants to observe a $P_{hit}$ value of 0 while the trial is still in progress when the actual $P_{hit}$ value is less than $0.1$ but not exactly 0.  

A total of $48$ unique $P_{hit}$ trajectories were used in the experiment; some trajectories were repeated for trials with different group sizes, group protocols, or shelter capacities, and participants were not informed when a $P_{hit}$ trajectory was repeated.

\textbf{Disaster Progress panel.} The computer interface (Fig \ref{fig:overview}B) provides each participant with current disaster and social information. Located at the top of the application window is a panel containing information on the progression of the disaster. The dark purple bar indicates the time progression and moves from left to right, indicating elapsed time. When the dark purple bar enters the red region, the disaster can hit or miss at any time. The light purple zone to the left of the red region represents the first 20 time steps of the trial, in which the disaster cannot hit or miss.

\textbf{Disaster Likelihood panel.} The graphical form in which probabilistic information is displayed has been shown to impact the accuracy of a decision informed by this information \cite{edwards}. 
{We choose to display disaster likelihood as a coarse-grained value which is relatively simple to comprehend and which reflects the limited amount of information available during a natural disaster.}
We represent disaster likelihood by the proportion of filled red circles; for example, in Fig \ref{fig:overview}B, {5} out of 10 circles are filled, indicating a disaster likelihood of {5} chances out of 10.

\textbf{Group Information panel.} The center portion of the interface is dedicated to social information and displays the evacuation decisions and ranks of other members in a participant's group. Each participant is represented by a unique avatar, which is accompanied by a number that indicates the participant's current rank. The avatar corresponding to the participant viewing the interface is displayed as a rectangular shape, while the other avatars are displayed as circles when the associated participants are at home and turn into triangles as the participants decide to evacuate. When the entire group evacuates, all members' avatars are replaced with images of shelters.

\textbf{Shelter panel.} The maximum shelter capacity in each trial is 5, 25, or 50 spaces. The shelter panel displays the number of available spaces in the shelter. Each available space is represented by a gray rectangle, which changes to a red rectangle when it becomes occupied.

\textbf{Evacuate button.} Participants click the Evacuate button when they decide to evacuate. In individual trials, if there is enough room in the shelter, participants immediately evacuate and occupy a shelter space once they have clicked the Evacuate button. In group trials, participants click the Evacuate button to ``vote'' for a group evacuation but do not necessarily immediately evacuate. The group does not evacuate until enough evacuation decisions (``votes'' for evacuation) have been made; the number of decisions required for group evacuation depends on the specific evacuation protocol. Once enough participants in a group have decided to evacuate, all group members evacuate at once.

At the beginning of the experiment, participants undergo a training session and perform 16 ``practice'' trials in order to familiarize themselves with the experimental setup. The data from the 16 practice trials were excluded from our analysis and modeling process. During the practice trials, a \textit{Leave Shelter} button was included, which was then disabled for the 144 main trials; during the main trials, participants were not allowed to return to their homes once they had evacuated. 

\begin{figure}[ht]
\centering
\includegraphics[scale=0.5]{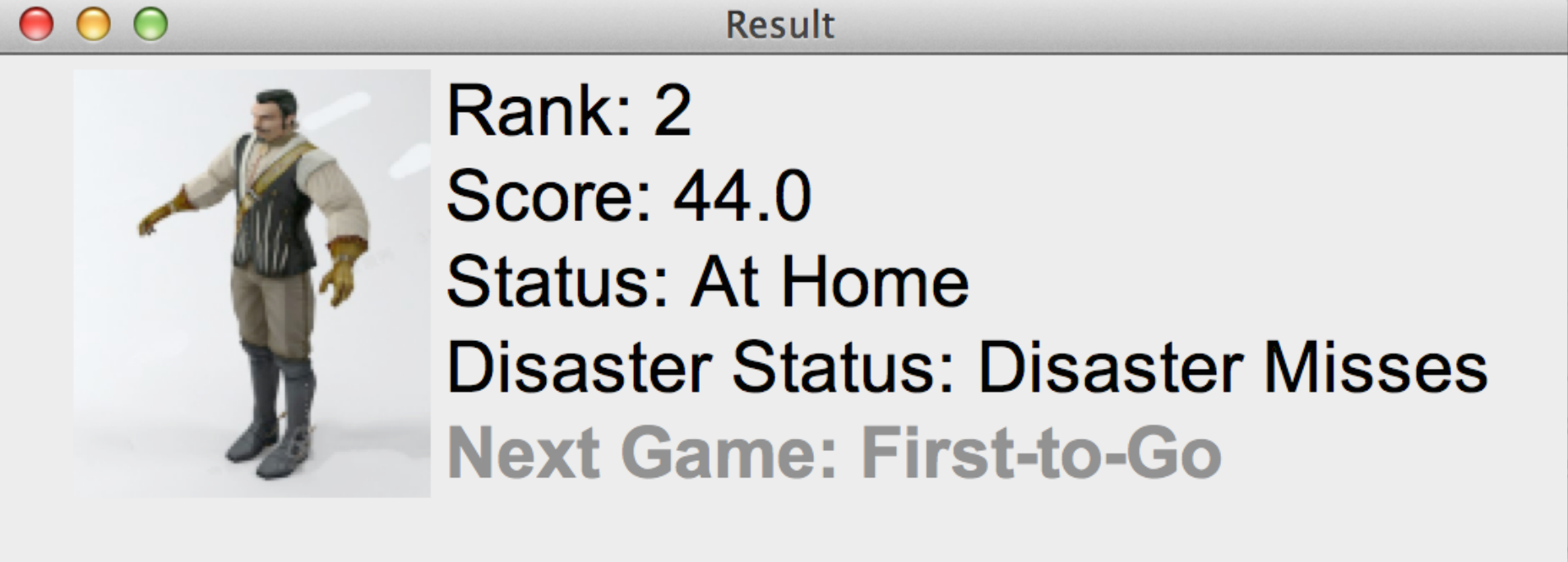}
\caption{\textbf{Summary window.} The summary window displayed after each trial shows the individual's current rank among 50 participants, current total score, personal evacuation status of the previous trial, the final status of the previous disaster, and the group protocol of the upcoming trial.}
\label{fig:summary}
\end{figure}

\textbf{Summary window.} After each trial, a summary window (Fig \ref{fig:summary}), is displayed; the summary window is unique to each participant and contains information on their rank, their total score, their decision in the previous trial, the final status of the disaster in the previous trial (i.e., whether it hit or missed), and the group protocol of the upcoming trial, allowing participants to review their prior outcomes and prepare for the next trial.

\textbf{Scores and rankings.} For each trial, individuals are staked 10 points, out of which a number of points are deducted based on the success of their actions in the trial (Table \ref{table:lossmat}). The loss matrix (top half of Table \ref{table:lossmat}) aims to reflect the risks inherent in evacuating in a real-world scenario. We chose to encode risk information in a loss matrix rather than display the outright incentives, since in a real-world disaster scenario, major concerns would be oriented around loss of life and property as opposed to any potential gains. For instance, more points are lost on average when the disaster hits than when the disaster misses, since individuals could potentially lose their homes, regardless of whether they evacuated, as well as their lives, had they remained at home. Furthermore, while zero points are deducted when a participant remains home during a disaster miss, a small number of points are deducted when a participant evacuates during a disaster miss, representing the loss of time and resources (such as gasoline) resulting from this incorrect decision.

The loss matrix is briefly shown to participants during a training session prior to the first trial but is not emphasized in order to minimize the influence of specific loss values on participants' decisions. Its values are fixed for all trials to allow greater variability with respect to other variables in the nested experimental design, namely disaster likelihood, shelter space, group size, and group protocol.

\textbf{Group sizes and protocols.} 
All trials are either individual or group trials, with the order chosen randomly prior to the experiment. In group trials, all participants evacuate collectively as 5-person groups (``group-5'') or as 25-person groups (``group-25''), depending on the trial. At the beginning of the experiment, participants are randomly assigned to groups, and group membership remains the same for all trials. 
That is, if two participants are group-mates in one trial, they remain group-mates in all trials with the same group size. All 50 participants take part in each of the 144 trials.

In group trials, the action of all groups is mandated by one of three specified protocols: first-to-go (FTG), majority-vote (MV), and last-to-go (LTG). In the first-to-go protocol, an entire group evacuates at once when the first group member decides to evacuate. In the majority-vote protocol, a whole group evacuates when more than half of the group members decide to evacuate. In the last-to-go protocol, a whole group will evacuate only when all group members decide to evacuate. As in individual trials, once a group evacuates, they cannot return home from the shelter.

Since clicking the Evacuate button need not coincide with evacuation in certain group scenarios, we differentiate between the \textit{decision} to evacuate and the \textit{evacuation} itself. A participant makes a decision to evacuate when they click the Evacuate button. This immediately results in evacuation in individual trials and in FTG trials, as well as in MV and LTG trials if the participant's decision is the final decision required to surpass the threshold specified by the protocol. Furthermore, it is also possible for a participant to evacuate without deciding to evacuate; when the threshold has been surpassed in FTG or MV trials, participants who have not made decisions to evacuate will evacuate with the rest of their group.

\section*{Empirical Results}
In this section, we characterize decision-making behavior based on our empirical observations and quantify the influence of key variables, including the disaster likelihood trajectory, group size, group evacuation protocol, and individual personality.

\begin{figure}[ht]
\centering
\hspace{-3.35cm}
\includegraphics[scale=1]{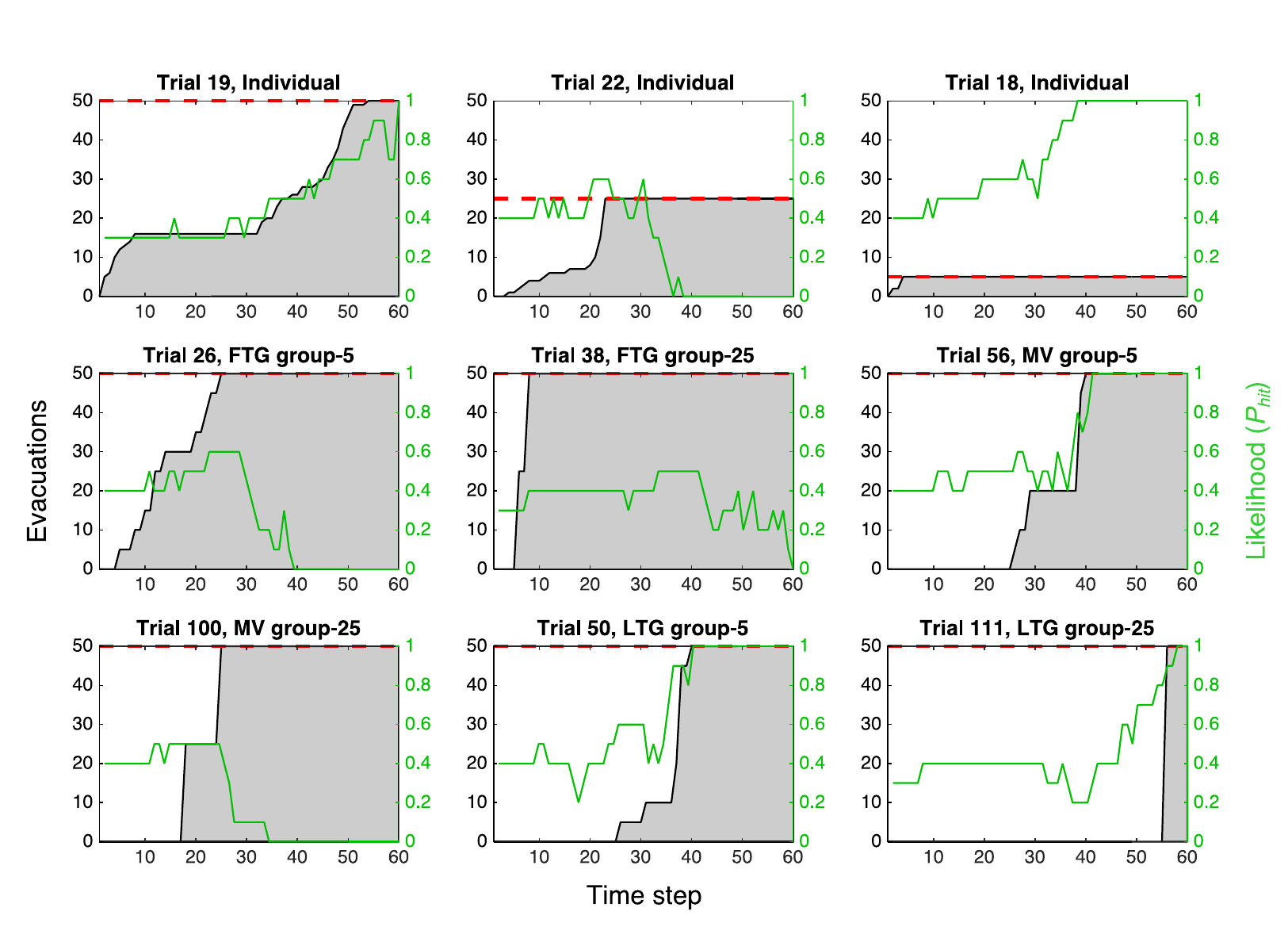}
\caption{\textbf{Observed evacuation behavior in nine example trials.} The cumulative evacuations (gray area) and disaster likelihood (green line) as a function of time step are plotted for individual trials with shelter spaces of 5, 25, and 50, and for FTG, MV, and LTG group trials with group sizes of 5 and 25. The maximum available shelter space is indicated by the dashed red line.}
\label{fig:0game}
\end{figure}

Examples of our observations are plotted in Fig \ref{fig:0game}. The top three panels show the cumulative evacuations (gray area) for three individual trials with initial shelter spaces of 50, 25, and 5 spaces (dashed red line), respectively. In the first panel, Trial 19, the trajectory of the disaster likelihood (green line) increases almost monotonically over the course of the trial with some small fluctuations. The number of evacuations increases quickly very early in the trial but remains constant while $P_{hit}$ remains near a value of $0.3$. Around time step $t = 30$, the $P_{hit}$ value begins to steadily increase, and the number of evacuations increases as well. All participants evacuate by time step $t = 50$.

In the second panel, Trial 22, the $P_{hit}$ value initially fluctuates between $0.4$ and $0.5$ up until $t = 20$, at which point it quickly increases from $0.4$ to $0.6$. A large wave of evacuations immediately follows this $P_{hit}$ increase, causing the shelter to fill up at $t = 22$. However, the disaster ultimately misses.

In the third panel, Trial 18, the shelter has a maximum capacity of 5 spaces and fills up at $t = 4$, even though $P_{hit}$ remains constant during the first 4 time steps of the trial. We observe that for all 5-space individual trials, participants typically evacuate very early in the trial, despite minimally varying $P_{hit}$ values.

The bottom two rows of Fig \ref{fig:0game} illustrate six example group trials. Evacuations typically occur earlier in a trial for groups following the FTG protocol than MV or LTG. For example, in the fifth panel, Trial 38, both groups evacuate by $t = 10$, even though $P_{hit}$ is relatively low, having increased from $0.3$ to $0.4$; the disaster ultimately misses in this trial. In contrast, in the last panel, Trial 111, both groups do not evacuate until $t = 55$, when $P_{hit}$ has climbed from a local minimum of $0.2$ to $0.8$ before the disaster ultimately strikes shortly after.

Out of 144 total trials, the data from 16 trials were excluded due to technical difficulties in recording data, for a total of 128 trials included in our analysis and modeling.
There were a total of 18 possible configurations of group size (individual, group-5, and group-25), protocol (FTG, LTG, and MV), and shelter capacity (5, 25, and 50 spaces), each represented by at least one trial in the experiment. Out of the 128 trials included in the analysis, 46 were individual trials, out of which there were 16 trials with 50 shelter spaces, 13 with 25 spaces, and 15 with 5 spaces. 82 trials were group trials, out of which 27 were FTG trials (20 group-5 trials, 7 group-25 trials), 29 were MV trials (15 group-5 trials, 14 group-25 trials), and 26 were LTG trials (13 each group-5 and group-25 trials). 

\begin{figure}[ht!]
\centering
\includegraphics[scale=0.75]{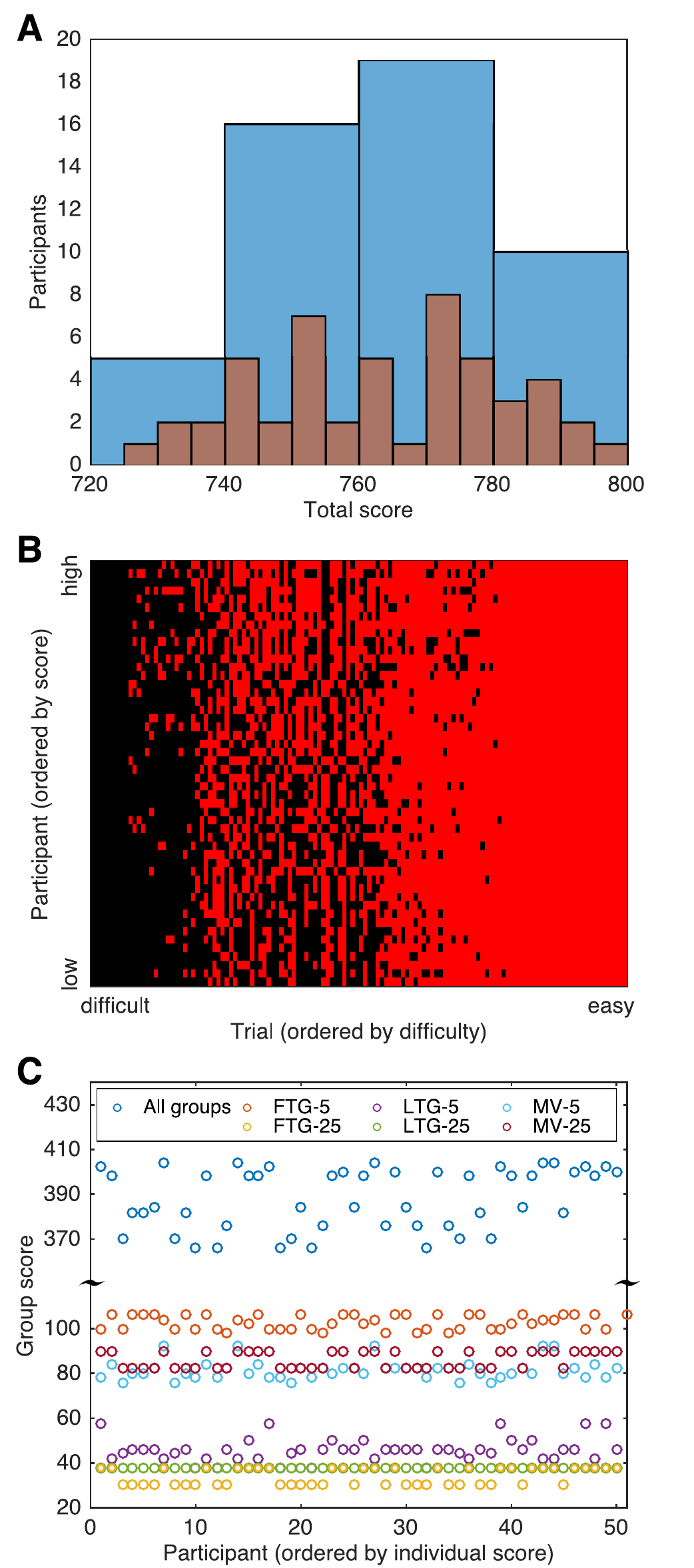}
\caption{\textbf{Distribution of scores.} Panel A is a histogram of total scores (blue) with orange bars displaying the score distribution at finer resolution. Panel B shows successful end states in red [(\textit{In Shelter}, \textit{Hit}); (\textit{At Home}, \textit{Miss})] and unsuccessful end states in black. Participants are sorted by the total score over all trials. Panel C plots the total score for each group size and protocol, as well as all group trials combined, for all participants, ordered by their total score over all individual trials. Scores from only the final 105 trials are plotted in C, due to a reassignment of groups after the first 23 trials.}
\label{fig:distcumu}
\end{figure}

\textbf{Participant rankings and scores.} 
Fig \ref{fig:distcumu}A plots the distribution of total scores for all participants, with orange bars showing the distribution at a finer resolution than the blue bars. The success of each participant's actions is summarized in Fig \ref{fig:distcumu}B. We quantify a participant's success by their total score achieved at the conclusion of the 128 trials. The two successful end states {[(\textit{In Shelter}, \textit{Hit}); (\textit{At Home}, \textit{Miss})]} are represented by red patches, while unsuccessful states {[(\textit{At Home}, \textit{Hit}); (\textit{In Shelter}, \textit{Miss})]} are represented by black patches. The participants are ordered by their total score, with the highest-scoring participant at the top. The 128 trials are ordered by difficulty, with the most difficult trial on the left. Here difficulty refers to the proportion of unsuccessful decisions to all decisions made in a trial. In the nine most difficult trials, none of the participants were successful, whereas all participants were successful in the 28 least difficult trials. 

Fig \ref{fig:distcumu}C shows that better performances in individual trials do not necessarily translate to better performances in group trials; there is no significant correlation between total scores in individual trials and total scores in group trials.
However, for scores totaled over all group trials, there are two distinct clusters, each containing 25 participants, separated by a gap in the distribution between score values 385 and 397. Each of these clusters corresponds to the one of the two groups of 25; one group generally performs better than the other group in group-25 trials. While every participant has the same score in LTG group-25 trials (green circles), the same group performs better than the other group in FTG (yellow circles) and MV trials (dark red circles).

Furthermore, the five participants who rank highest in individual trials are part of the higher-scoring group of 25. While both groups perform equally well on average in LTG group-25 trials, in the higher-scoring group, we observe a statistically significant positive correlation (Spearman rank correlation coefficient = 0.6238, $p < 0.01$) between the order of evacuation decision and rank, i.e., higher-scoring participants tend to decide to evacuate earlier than lower-scoring participants. Additionally, we observe a similar pattern of behavior in the same group for MV group-25 trials, with a Spearman coefficient of $0.5769$, $p < 0.05$. In the lower-scoring group, we do not observe a statistically significant correlation between evacuation decision order and rank in either LTG or MV trials. This suggests that in successful groups, lower-scoring participants may be influenced by group members who are ranked more highly, and they may wait to act until higher-ranking group members have decided to evacuate. 

\begin{figure}[ht!]
\centering
\includegraphics[scale=1]{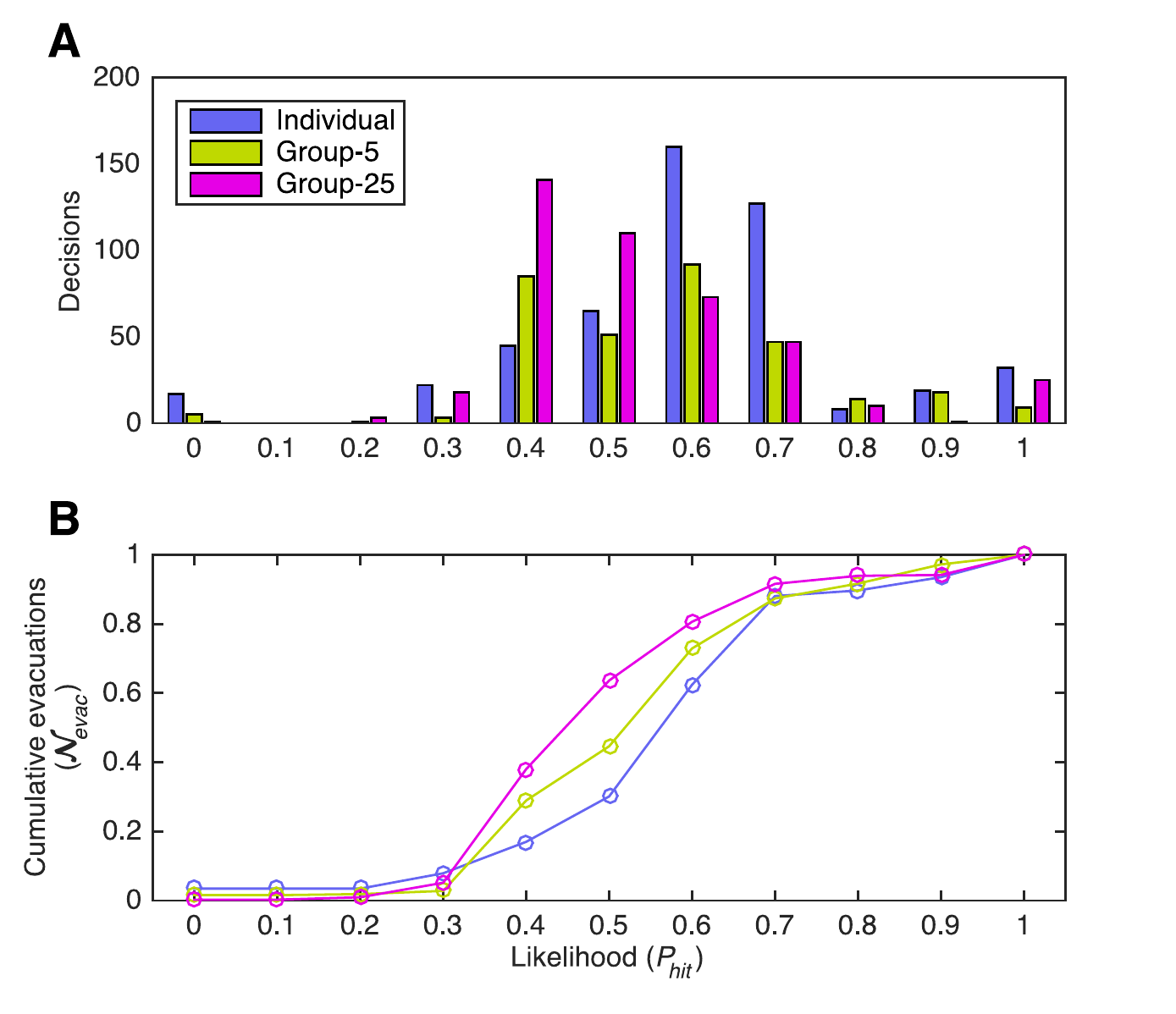}
\caption{\textbf{Histogram and distribution of evacuation decisions as a function of $P_{hit}$.} Panel A compares the numbers of evacuation decisions as a function of $P_{hit}$ between individual trials (blue) and group trials (green, 5-person groups; magenta, 25-person groups). Panel B compares the normalized cumulative number of decisions $\mathcal{N}_{dec}$ for individual and group trials. All trials shown have a maximum of 50 shelter spaces.} \label{fig:phit}
\end{figure}

\textbf{Disaster likelihood.}
Decisions to evacuate were strongly influenced by the disaster likelihood, as illustrated in Fig \ref{fig:phit}. Fig \ref{fig:phit}A plots a histogram of evacuation decisions as a function of the displayed (i.e., rounded) $P_{hit}$ value at the time of the decision, and Fig \ref{fig:phit}B plots the corresponding cumulative normalized evacuation decisions $\mathcal{N}_{dec}$. Results are plotted separately for individual trials, group-5 trials, and group-25 trials; in all trials shown, the maximum shelter capacity is 50 spaces. Note that even though the trials end when $P_{hit} = 1$, a slight reaction time or communication delay between the server and the individual interfaces can result in participants viewing a $P_{hit} = 1$ value on the last time step of a trial and making an evacuation decision on this time step. Evacuation decisions made at $P_{hit} = 1$ are included in this figure.

The distribution of decisions in individual trials (blue bars in Fig \ref{fig:phit}A) has a local maximum at $P_{hit}=0.6$.
Note the relatively low number of decisions at $P_{hit} = 0.8$ and $0.9$. While the probability of a disaster hit is very close to certain in these cases, most participants likely have already decided to evacuate at smaller values of $P_{hit}$. 
Compared to individual decisions, decisions made in group trials occur on average at lower values of $P_{hit}$, and the larger the group, the lower the average value of $P_{hit}$. {The mean $P_{hit}$ value for individual decisions is 0.65 with standard deviation 0.19, compared with $0.62 \pm 0.17$ for groups of 5, and $0.59 \pm 0.16$ for groups of 25.}
For group-5 trials (green bars in A), the distribution has two local maxima at $P_{hit} = 0.4$ and $P_{hit} = 0.6$, and for group-25 trials (magenta bars), the distribution has a maximum at $P_{hit} = 0.4$. The plots of cumulative normalized decisions ($\mathcal{N}_{dec}$) in Fig \ref{fig:phit}B illustrate that on average, decisions to evacuate are made at decreasing disaster likelihood $P_{hit}$ as the group size increases.
This suggests that participants follow different strategies when making evacuation decisions as part of a group. Furthermore, the individual and group trial results are sufficiently different such that the group behavior cannot be explained as a biased sample of the individual trials. {We determine the statistical significance of the absolute difference in means between each pair of distributions by performing a permutation test, which does not assume that the two distributions to be compared are normal.
The permutation test is performed by randomly reshuffling the data labels of two distributions to generate two new distributions equal in size to the original pair, calculating the absolute difference in their means, and repeating for a total of 5000 times; the $p$-value of the difference in means is calculated as the percentage of differences greater than the observed difference. Individual behavior is statistically different from group-5 at the 95\% confidence level ($p = 0.05$) and group-25 ($p < 0.001$) behavior, and group-5 is statistically different from group-25 ($p = 0.003$).}

\begin{figure}[ht]
\centering
\hspace{-0.8cm}
\includegraphics[scale=1]{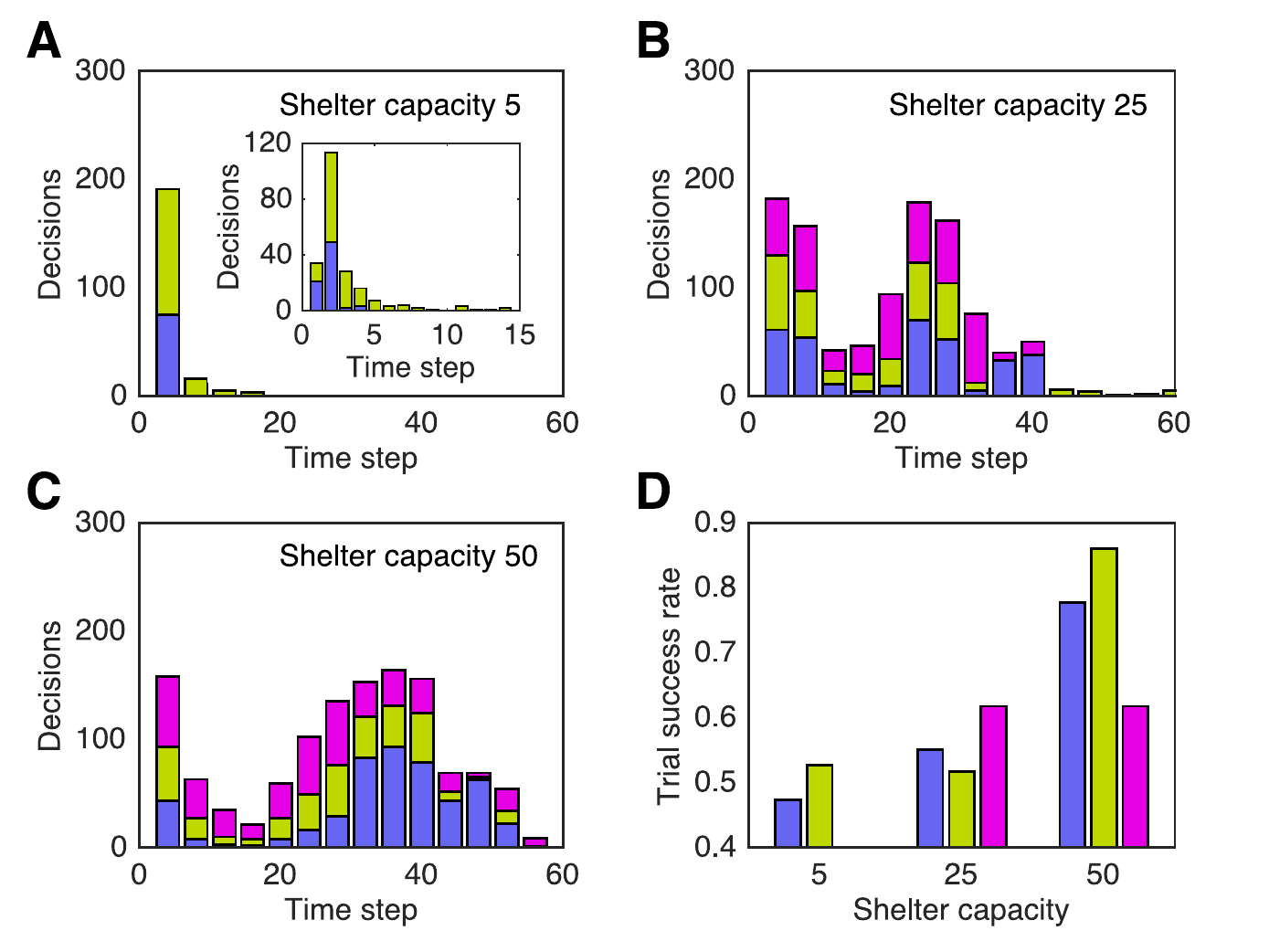}
\caption{\textbf{Comparison of shelter sizes.} Panel A (B, C) plots a histogram of evacuation decisions as a function of experiment time for trials with 5 (25, 50) shelter spaces. Bars are separated into sections representing group-25 trials (magenta), group-5 trials (green), and individual trials (blue). All plots are shown on the same scale; panel A contains an inset with the same plot at higher temporal resolution. Panel A demonstrates that participants tend to decide to evacuate at early time steps when shelter space is extremely limited. Panel D plots the trial success rate as a function of shelter space.}
\label{fig:spaceTime}
\end{figure}

\textbf{Shelter space.}
Participants generally decide to evacuate at earlier time steps when shelter space is limited. Figs \ref{fig:spaceTime}ABC display histograms of evacuation decisions as a function of time for 5, 25, and 50-space trials, respectively. Fig \ref{fig:spaceTime}D plots a histogram of the trial success rate as a function of shelter space. Participants decide to evacuate at earlier time steps when shelter space is limited and have lower success rates for smaller shelter capacities. In 5-space trials (Fig \ref{fig:spaceTime}A), all decisions occur before time step $t = 15$, even though trials are guaranteed to continue beyond $t=20$. The inset shows that most decisions occur at $t=2$. This indicates that many participants make evacuation decisions in 5-space trials as early as possible in order to avoid being locked out of the shelter.

In 25-space trials (Fig \ref{fig:spaceTime}B), there are two local maxima located around time steps $t=4$ and $t = 24$, which we refer to as the ``early-decision peak'' and the ``late-decision peak''.
For 50-space trials (Fig \ref{fig:spaceTime}C), there are also two local maxima; the early-decision peak also occurs at $t=4$, while the late-decision peak is located at $t = 36$. 
Unexpectedly, the maximum values of both early-decision peaks are comparable to their respective late-decision peaks. Although trials are guaranteed to continue past $t = 20$, for all shelter capacities, a significant number of decisions occur at early time steps even when the shelter space is sufficiently large to accommodate all participants.

Participants generally also have a higher success rate when shelter space is not limited (Fig \ref{fig:spaceTime}D). Groups of 5 have the highest success rate of 0.860 for 50-space trials, compared with 0.776 for individuals and 0.618 for groups of 25. For 25-person groups, the trial success rate is equal for both 25-space and 50-space trials (0.618).

\begin{figure}[ht]
\centering
\hspace{-0.8cm}
\includegraphics[scale=1]{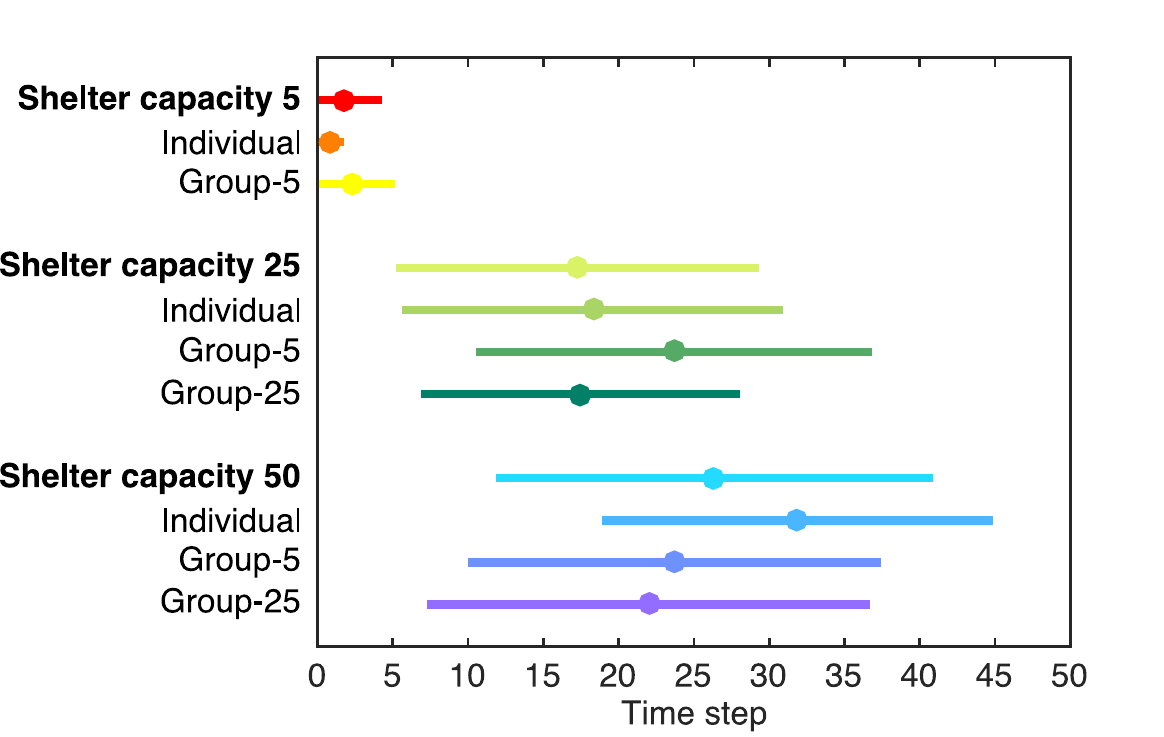}
\caption{\textbf{Comparison of mean and standard deviation for evacuation decision time step.} The average  time step at which evacuation decisions were made and the corresponding standard deviation is plotted for trials with shelter capacity 5, 25, and 50 (individual and group trials combined), and the respective individual, group-5, and group-25 trials.}
\label{fig:spaceTimemeans}
\end{figure}

Fig \ref{fig:spaceTimemeans} compares the mean and standard deviation of evacuation decision times for each shelter capacity and group size. {The average evacuation decision time for individual and group trials combined increases as shelter capacity increases; the mean decision time is 1.8 time steps for 5-space trials, 17.3 for 25-space, and 26.3 for 50-space trials.}
Table \ref{table:spacetimettest} lists the $p$-values for the permutation test results comparing decision times for different shelter capacities under a fixed group size, and different group sizes under a fixed shelter capacity. We find significant differences in decision making between most scenarios; for a fixed group size, the decision times for different shelter capacities have statistically different means with $p < 0.001$. In contrast, for trials with 25 shelter spaces, the distribution of decision times in individual trials is not statistically different under the permutation test from that of group-25 trials with a confidence level of at least $95\%$, and likewise with group-5 trials compared to group-25 trials. Furthermore, for trials with 50 shelter spaces, groups of 5 do not decide to evacuate at significantly different times than groups of 25.

\begin{table}[t!] 
\caption{\textbf{Permutation Test Results Comparing Different Group Sizes and Shelter Capacities.}}
\begin{tabular}[t!]{|c|c|c|c|} 
\hline
Group Size & Shelter Capacity 1 & Shelter Capacity 2 & $p$ \\
\hline
\multirow{3}{*}{Individual} & 5 & 25 & $\mathbf{<0.001}$ \\
& 5 & 50 & $\mathbf{< 0.001}$ \\
& 25 & 50 & $\mathbf{<0.001}$ \\ 
\hline
\multirow{3}{*}{5} & 5 & 25 & $\mathbf{<0.001}$ \\
& 5 & 50 & $\mathbf{< 0.001}$ \\
& 25 & 50 & $ \mathbf{<0.001}$ \\
\hline
25 & 25 & 50 & $\mathbf{<0.001}$ \\
\hline
\end{tabular}
\vspace{10pt}

\begin{tabular}[t!]{|c|c|c|c|}
\hline
Shelter Capacity & Group Size 1 & Group Size 2 & $p$ \\
\hline
5 & Individual & 5  & $\mathbf{<0.001}$ \\
\hline
\multirow{3}{*}{25} & Individual & 5  & $\mathbf{0.015}$ \\
& Individual & 25 & $0.31$ \\
& 5 & 25  & $ 0.065$ \\ 
\hline
\multirow{3}{*}{50} & Individual & 5  & $\mathbf{<0.001}$ \\
& Individual & 25 & $\mathbf{<0.001}$ \\
& 5 & 25  & $0.10$ \\ 
\hline
\end{tabular}
\label{table:spacetimettest}
\caption*{Permutation test results comparing evacuation decision times for different shelter capacities with group size fixed, and comparing group sizes with shelter capacity fixed. The permutation test is performed by randomly reshuffling the labels of the data points in the two distributions to be compared, repeating to create 5000 different pairs of distributions, determining the absolute difference in the means between each pair of distributions, and calculating the $p$-value as the percentage of differences greater than the observed difference. Bold values indicate a rejection of the null hypothesis of equal means at the 95\% confidence level.}
\end{table}

\begin{figure}[!ht]
\centering
\hspace{-0.8cm}
\includegraphics[scale=1]{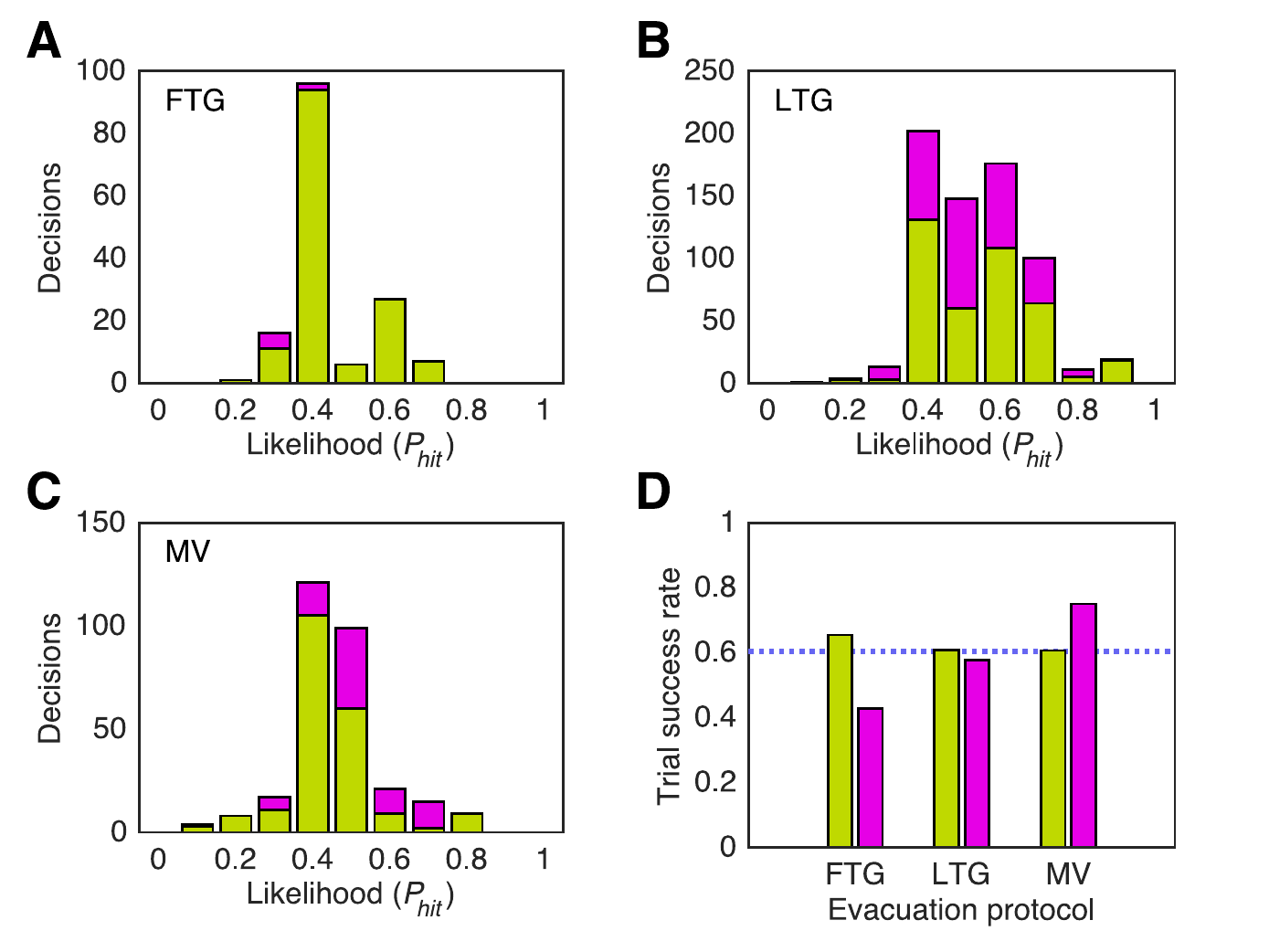}
\caption{\textbf{Comparison of group protocols.} Histograms of decisions as a function of disaster likelihood $P_{hit}$ are plotted in panels A, B, and C for the first-to-go, majority-vote, and last-to-go protocols, respectively, with bars divided by group-5 trials (green) and group-25 trials (magenta). Many decisions are made with $P_{hit}=0.4$ in the FTG trials as the decisions tend to be made in the beginning of trials. Panel D compares the trial success rate across group protocols. The dashed blue line represents the trial success rate for individual trials.}
\label{fig:prinProb}
\end{figure}

\textbf{Group protocols.} We compare the three group evacuation protocols, first-to-go, majority-vote, and last-to-go, in Fig \ref{fig:prinProb}. Figs \ref{fig:prinProb}ABC plot histograms of evacuation decisions as a function of $P_{hit}$ for each protocol. Since only one individual per group can make an evacuation decision in FTG trials, there are much lower numbers of decisions for group-25 trials, as shown in Fig \ref{fig:prinProb}A. For instance, there are 94 decisions made at $P_{hit} = 0.4$ in group-5 trials but only 2 decisions made at $P_{hit} = 0.4$ in group-25 trials. Furthermore, the majority of decisions made in FTG trials are made at $P_{hit} = 0.4$, since decisions in FTG trials tend to be made near the start of trials. 29\% of decisions in FTG trials are made at $t=1$, compared with 7\% for LTG and 8\% for MV.

Fig \ref{fig:prinProb}D compares the trial success rates for each protocol. The success rate for group-5 is similar for the three protocols, with the highest success rate occurring for FTG (0.655). The group-25 success rate is lower than that of group-5 for all protocols except MV. The group-25 MV success rate (0.75) is the highest of all protocols and group sizes. The success rate for individual trials (0.604) is also displayed (dashed blue line); it is approximately equal to that of the LTG and MV group-5 trials (0.608 and 0.607, respectively). The individual success rate is higher than that of all group-25 trials except for MV.

\begin{figure}[!ht]
\centering
\hspace{-0.8cm}
\includegraphics[scale=1]{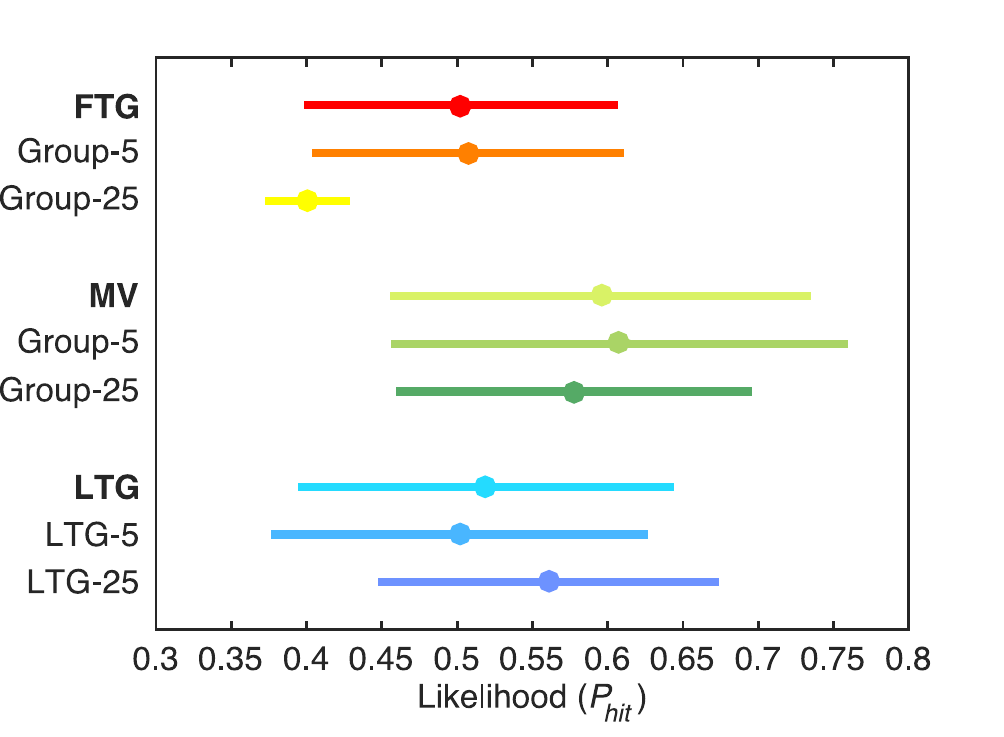}
\caption{\textbf{Comparison of mean and standard deviation for disaster likelihood $P_{hit}$ at the time of decision in group trials.} The average value of $P_{hit}$ at the time of decision and the corresponding standard deviation is plotted for FTG, MV, and LTG trials (all group sizes combined), and the respective group-5 and group-25 trials.}
\label{fig:prinProbmeans}
\end{figure}

Fig \ref{fig:prinProbmeans} compares the mean and standard deviation of evacuation decision times between different protocols and group sizes. The permutation test $p$-values are listed in Table \ref{table:prinprobttest}; distributions of decision times for different group sizes are compared for fixed group protocol, and different group protocols are compared for fixed group size. The null hypothesis of equal means in two compared populations is rejected at a 95\% confidence level in most cases except for FTG group-5 and LTG group-5, and for LTG group-25 and MV group-25.

\begin{table}[!hb]
\caption{\textbf{Permutation Test Results Comparing Different Group Sizes and Group Protocols.}}
\begin{tabular}[t!]{|c|c|c|c|} 
\hline
Group Protocol & Group Size 1 & Group Size 2 & $p$ \\
\hline
FTG & 5 & 25  &   $\mathbf{0.007}$ \\
LTG & 5 & 25  &  $\mathbf{<0.001}$ \\
MV & 5 & 25   & $\mathbf{0.004}$ \\ 
\hline
\end{tabular}
\vspace{10pt}

\begin{tabular}[t!]{|c|c|c|c|} 
\hline
Group Size & Group Protocol 1 & Group Protocol 2 & $p$ \\
\hline
\multirow{3}{*}{5} & FTG & LTG  & $0.66$ \\
& FTG & MV& $\mathbf{< 0.001}$ \\
& LTG & MV & $ \mathbf{<0.001}$ \\ 
 
\hline
\multirow{3}{*}{25} & FTG & LTG & $\mathbf{<0.001}$ \\
& FTG & MV  & $\mathbf{< 0.001}$ \\
& LTG & MV  & $ 0.24$ \\  
\hline
\end{tabular}
\label{table:prinprobttest}
\caption*{Permutation test results comparing evacuation decision times for different group sizes with group protocol fixed, and comparing group protocols with group size fixed. The permutation test is performed by randomly reshuffling the labels of the data points in the two distributions to be compared, repeating to create 5000 different pairs of distributions, determining the absolute difference in the means between each pair of distributions, and calculating the $p$-value as the percentage of differences greater than the observed difference. Bold values indicate a rejection of the null hypothesis of equal means at the 95\% confidence level.}
\end{table}

\begin{figure*}[ht!]
\centering
 \hspace*{-.18\textwidth}
\includegraphics[scale=0.95]{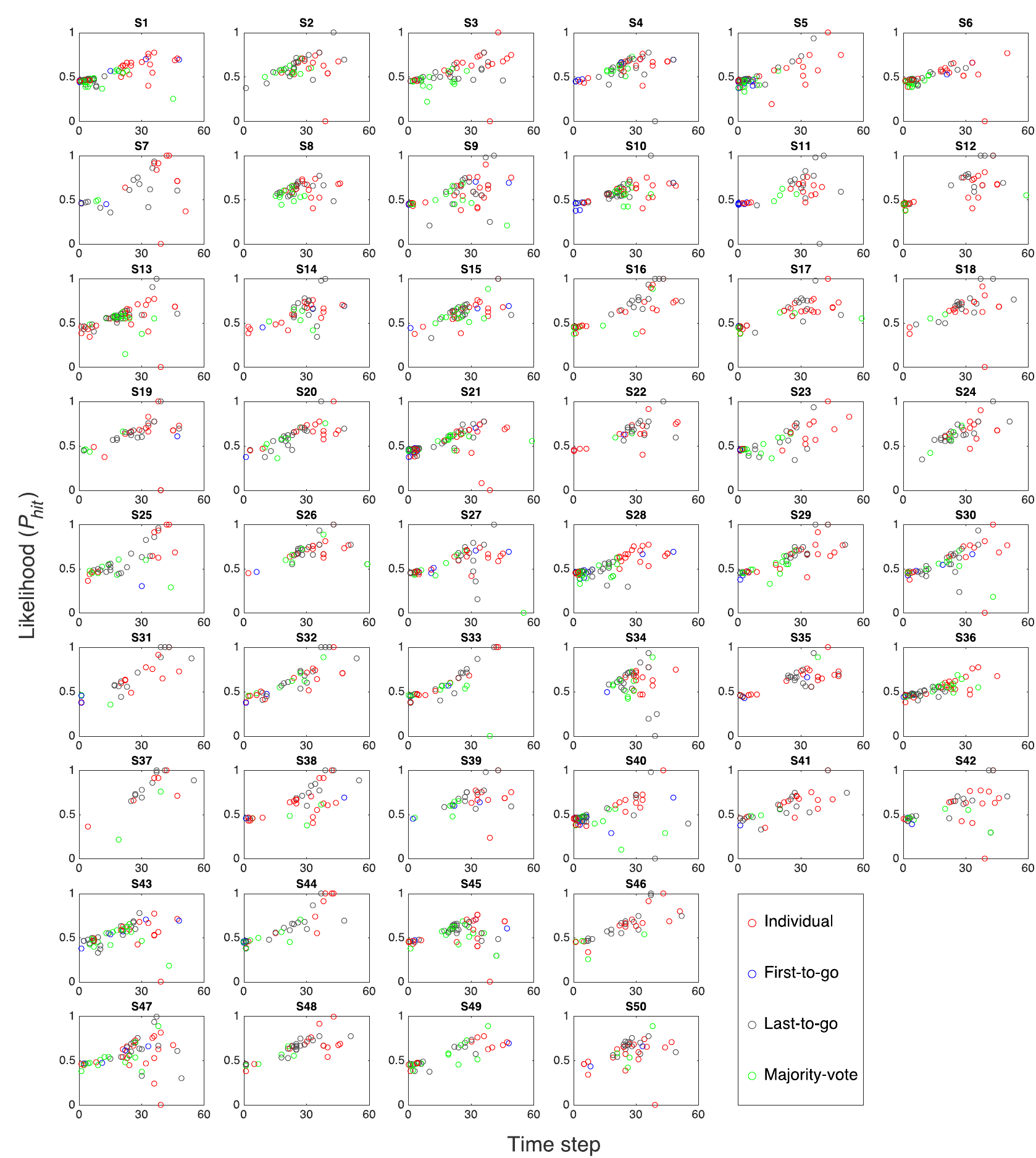}
\caption{\textbf{Individual differences in decision making.} The time and $P_{hit}$ (disaster likelihood) at each evacuation decision for each subject. Every evacuation decision is represented by a circle, with red denoting a decision in an individual trial, blue denoting FTG, black denoting MV, and green denoting LTG. Subplots correspond to individual subjects and are sorted according to final score, with the highest-scoring subject (S1) in the upper left corner and the lowest-scoring subject (S50) in the lower right corner. These plots demonstrate high variability in decision-making behavior across all subjects.}
\label{fig:phitvstime}
\end{figure*}

\textbf{Individual heterogeneity.}
We observe significant differences in behavior for different individuals. Fig \ref{fig:phitvstime} plots the time step and $P_{hit}$ of each evacuation decision for all participants; each subplot displays the decisions, represented by circles, of an individual participant. Subplots are ordered by performance, with the highest-scoring subject (S1) in the upper left corner and the lowest-scoring subject (S50) in the lower right corner.

We observe high variability in the distribution of decisions across all participants. For instance, S1's decisions in MV trials are largely clustered around $P_{hit} = 0.5$ and mostly occur at time steps less than 30, whereas S50's MV decisions are largely clustered near time step 30. S1's LTG decisions also occur relatively early in their respective trials compared to S50's LTG decisions.

Some participants' distributions of decisions have low variance in $P_{hit}$, such as S6 and S36, while others have low variance in time, such as S8 and S34. However, there is no correlation between variance in $P_{hit}$ or time and the participants' ranks. 
One participant with relatively anomalous behavior is S37; aside from a few instances, S37 typically makes decisions at relatively high values of $P_{hit}$ (0.6 or greater) and at relatively late time steps (30 or later).  

For most participants, decisions in individual trials are on average made at later time steps than decisions made during group trials. FTG and MV decisions are also typically made earlier than LTG decisions. Despite the wide range of individual variability illustrated in Fig \ref{fig:phitvstime}, coarse metrics that may be extracted from the individual patterns, such as an individual's average or variance in time step and/or $P_{hit}$ value for the evacuation decision, are not correlated significantly with individual rank. In the next section, we develop an artificial neural network that predicts individual and group decision making, which includes an additional node to capture individual personality factors from the data.

\section*{Modeling Decision Making}

\begin{figure}[ht!]
\centering
\hspace{-3cm}
\includegraphics[scale=1]{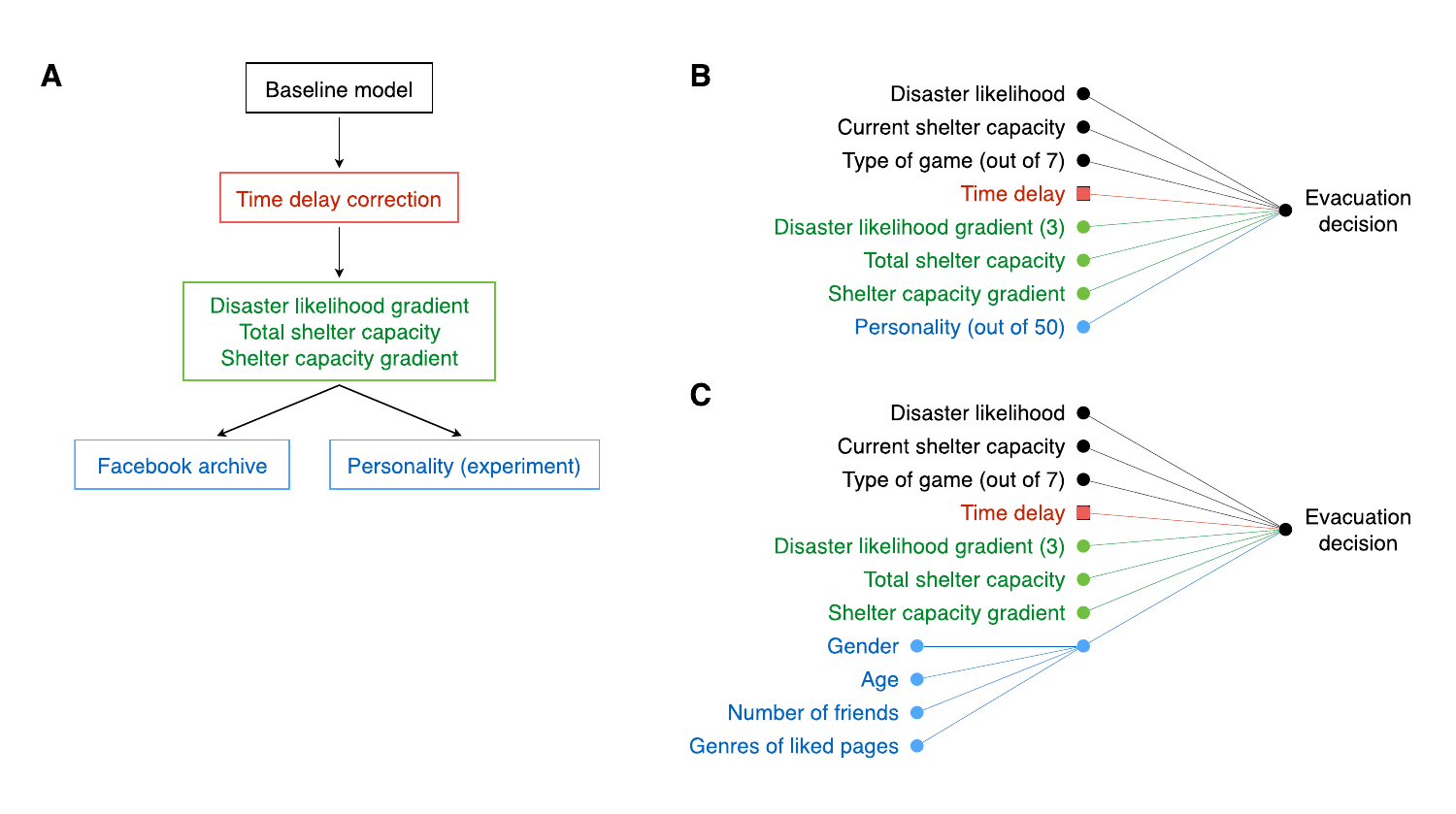}
\caption{\textbf{Model development flowchart and schematic of artificial neural networks.} Panel A describes the development of neural network models to predict evacuation behavior. We begin with a baseline model that takes as input disaster likelihood, current shelter capacity, and the type of trial (i.e., one out of 7 combinations of group size and protocol), which are represented by the black circles in the network schematics in panels B and C. To improve on this baseline model, we perform a time delay correction (red square). We then add input nodes for the disaster likelihood gradient (using data from three time steps), total shelter capacity, and shelter capacity gradient (green circles). We finally add an input node representing individual personality; the model in panel B includes a personality node (blue circle) that determines a unique personality factor (one per participant, out of 50) by training on the data, while the model in panel C uses Facebook likes and demographic information to generate the personality factor, which is incorporated as an input node to the neural network.}
\label{fig:nnflow}
\end{figure}

In this section, we develop a progressive series of artificial neural network models for individual and group decision making based on empirical observations and social media use. Fig \ref{fig:nnflow}A is a schematic illustration of the steps in our model development. We begin with a baseline model isolating key parameters of the experiment. We next augment the model with additional features extracted from the data to improve the overall accuracy. These include reaction time delay as well as gradients in the disaster likelihood and available shelter space. The final stage incorporates individual differences in experimental behavior, thereby inferring significant distinguishing characteristics (Fig \ref{fig:phitvstime}) that could not be determined using simple metrics such as the average or variance of aspects of individual behavior. Alternatively, we substitute representation of individual differences in the experiment with an additional network layer derived from social media use, which leads to quantitatively similar results. 

\textbf{Data balancing.} 
From the experimental data, we construct a set of positive and negative data points in non-sequential order; some data is then withheld for training, and the artificial neural network is trained on the remaining data.
The training and test data are assembled from the experimental data for all participants in all trials at every time step where the individual either decided to evacuate (a positive data point) or did not decide to evacuate, but could have (a negative data point). 
In group trials, positive data points occur when the individual is at home and decides to evacuate, whether or not their decision leads to a group evacuation at the particular time step. Negative data points occur when the individual is at home and does not decide to evacuate, but could have (i.e., there is sufficient space available in the shelter).  

Each individual can make at most one evacuation decision in each trial. In comparison, there are many time steps in which they remain at home and do not decide. As a result, uniform data sampling leads to a disproportionate number of negative points. To resolve this imbalance, we down-sample the negative points. To isolate the time leading up to a decision, we restrict our samples to lie within a window of width $t_w$ containing at least one positive point. For example, if a participant decides to evacuate at time $t$, we include in our training set negative data points between and including time steps $t-t_w$ and $t-1$. Here we set $t_w = 10$. Since this still results in excess negative data points, we up-sample the positive data points by doubling their weights.

\textbf{Baseline decision model.} We build our model incrementally as illustrated schematically in Fig \ref{fig:nnflow}A. We begin with a baseline logistic regression model that depends on key experimental variables: the disaster likelihood, the available shelter capacity, the group size, and the evacuation protocol. These inputs constitute the starting point for both of the final models (Fig \ref{fig:nnflow}B and C), which differ only in how individual differences are incorporated in the final step.

The baseline model takes vectors generated from the key variables as input and in turn outputs the probability of an evacuation decision,
\begin{equation}
\mathcal{P}_{dec}(n,t) = f(\textbf{w} \cdot \textbf{x}_{(n,t)} +b),
\label{eq:pdec}
\end{equation}
where, for trial $n$ and time step $t$, $\textbf{x}_{(n,t)}$ is generated by concatenating all input variables into a vector; \textbf{w} is a vector containing the weights of each of the terms of \textbf{x}; and $b$ is the bias term. The values of \textbf{w} and $b$ are determined by training on the data set. The function $f$ is usually defined as a continuous nonlinear function; we choose the sigmoidal function,

\begin{equation}
f(z)=\frac{1}{1+e^{-z}}. \label{eq:sigmoid}
\end{equation}

A 9-dimensional input vector $\textbf{x}_{(n,t)}$ is generated for each time step $t$ in each trial $n$ in the data set. The first two dimensions of $\textbf{x}_{(n,t)}$ correspond to the disaster likelihood $P_{hit}(n,t)$ and the number of available shelter spaces at time $t$ in trial $n$, respectively. The last 7 dimensions of $\textbf{x}_{(n,t)}$ represent the type of trial, i.e., the evacuation protocol and the group size, using ``1-out-of-$N$'' coding. That is, for each input vector $\textbf{x}_{(n,t)}$, only one of the last 7 dimensions has a value of 1, and all other dimensions have value 0. The 7 dimensions correspond to 7 possible combinations of group size and protocol: individual, FTG group-5, FTG group-25, MV group-5, MV group-25, LTG group-5, and LTG group-25, respectively. 

For example, trial 19 (shown in the first panel of Fig \ref{fig:0game}) is an individual trial. At $t = 34$, the disaster likelihood is 0.5. There are 31 spaces available in the shelter and 31 active participants. On that time step, one person decided to evacuate while the other 30 did not decide to evacuate. The input vector $\textbf{x}_{(n,t)}$ at time step 34 is then  $\textbf{x}_{(n=19,t=34)} = [0.5, 31,1, 0, 0, 0, 0, 0, 0]$. This vector contributes 31 data points, with 1 labeled as `1' and 30 labeled as `0', in our training dataset, as 31 people are still active in the trial at that time. These data points are used to train the values of \textbf{w} and $b$.

\begin{figure}[ht]
\centering
\includegraphics[scale=1]{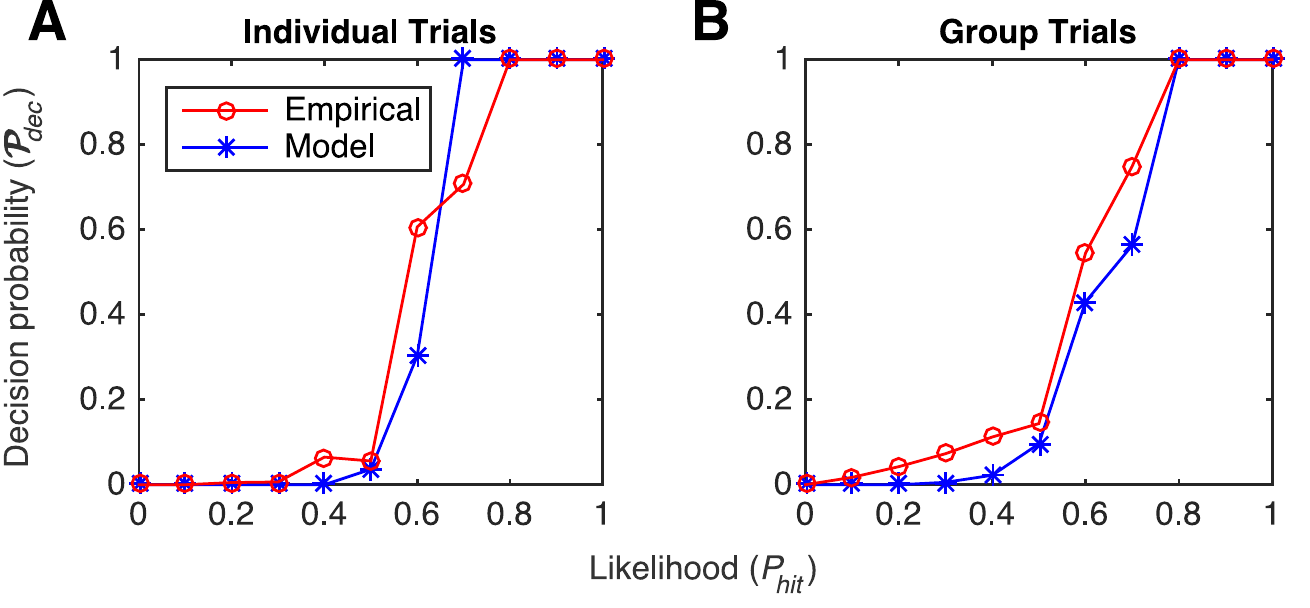}
\caption{\textbf{Logistic regression fit results.} The decision probability $\mathcal{P}_{dec}$ is plotted as a function of $P_{hit}$ for individual trials (A) and group trials (B). The red curve represents the empirical value and the blue curve represents the model.} 
\label{fig:logis}
\end{figure}

Fig \ref{fig:logis} compares the baseline model with the empirical decision probability as a function of $P_{hit}$ for individual and group trials. 
The model probability curve is generated by averaging, for each value of $P_{hit}$, all $\mathcal{P}_{dec}$ values output by the model when the associated input vector from the balanced training set contains that particular $P_{hit}$. The empirical probability curve is calculated by dividing the number of evacuation decisions (i.e., the number of positive data points) by the total number of data points (i.e., the sum of the numbers of positive and negative data points). The model curve is a fit to the empirical curve; both curves are generated from the same balanced (i.e., down-sampled and up-sampled as described previously) data set encompassing all 128 trials.

We observe that the model closely follows the data, indicating that the baseline logistic model accurately captures this average collective behavior observed experimentally. The empirical data plotted in Fig \ref{fig:logis} is taken from the balanced data set and includes both positive and negative data points, thus differing from the cumulative normalized evacuations $\mathcal{N}_{dec}$ in Fig \ref{fig:phit}B, which shows the empirical cumulative distribution of positive data points only (i.e., observed evacuation decisions), normalized across the entire experimental data set.

\begin{figure}[ht]
\centering
\includegraphics[scale=1]{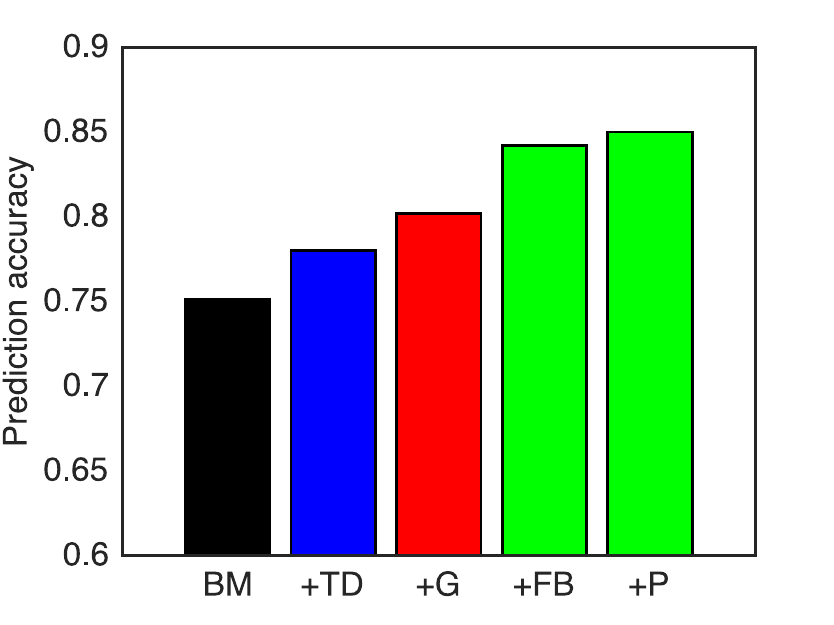}
\caption{\textbf{Prediction accuracy for baseline and enhanced models.} The prediction accuracy of the baseline logistic regression model (BM) is improved with addition of a time delay correction (+TD) and gradients in likelihood as well as gradient and total initial shelter space (+G). Furthermore, personality factors obtained either from Facebook data (+FB) or experimental data (+P) further increase the accuracy by accounting for individual differences. The evacuation prediction accuracy for the two complete models (+FB and +P) are comparable, with +P performing slightly better.}
\label{fig:measure}
\end{figure}

To evaluate the accuracy of the model (Fig \ref{fig:measure}) in predicting evacuation decisions, the entire balanced data set is first divided into a training set, validation set, and test set. The test set contains the balanced data points from 17 trials between trials 81 and 100. {The training set contains 80\% of the balanced data points from the remaining 111 trials, while the validation set comprises the other 20\% of the balanced data; the data points are selected randomly for each set. The 80-20 split between training and validation is commonly used in machine learning \cite{murphy}}. The training set is used to fit the parameters \textbf{w} and $b$ in Eq \eqref{eq:pdec}. The validation set is then used to convert the continuous decision probability $\mathcal{P}_{dec}$ into a binary decision. To accomplish this, we introduce a hyperparameter representing a decision boundary or threshold: a participant is predicted to make a decision to evacuate when the model outputs a probability higher than the decision boundary.
This hyperparameter is determined by selecting the decision boundary which best fits the data in the validation set.

\begin{figure*}[ht] 
\centering 
\hspace{-5.8cm}
\includegraphics[scale=1]{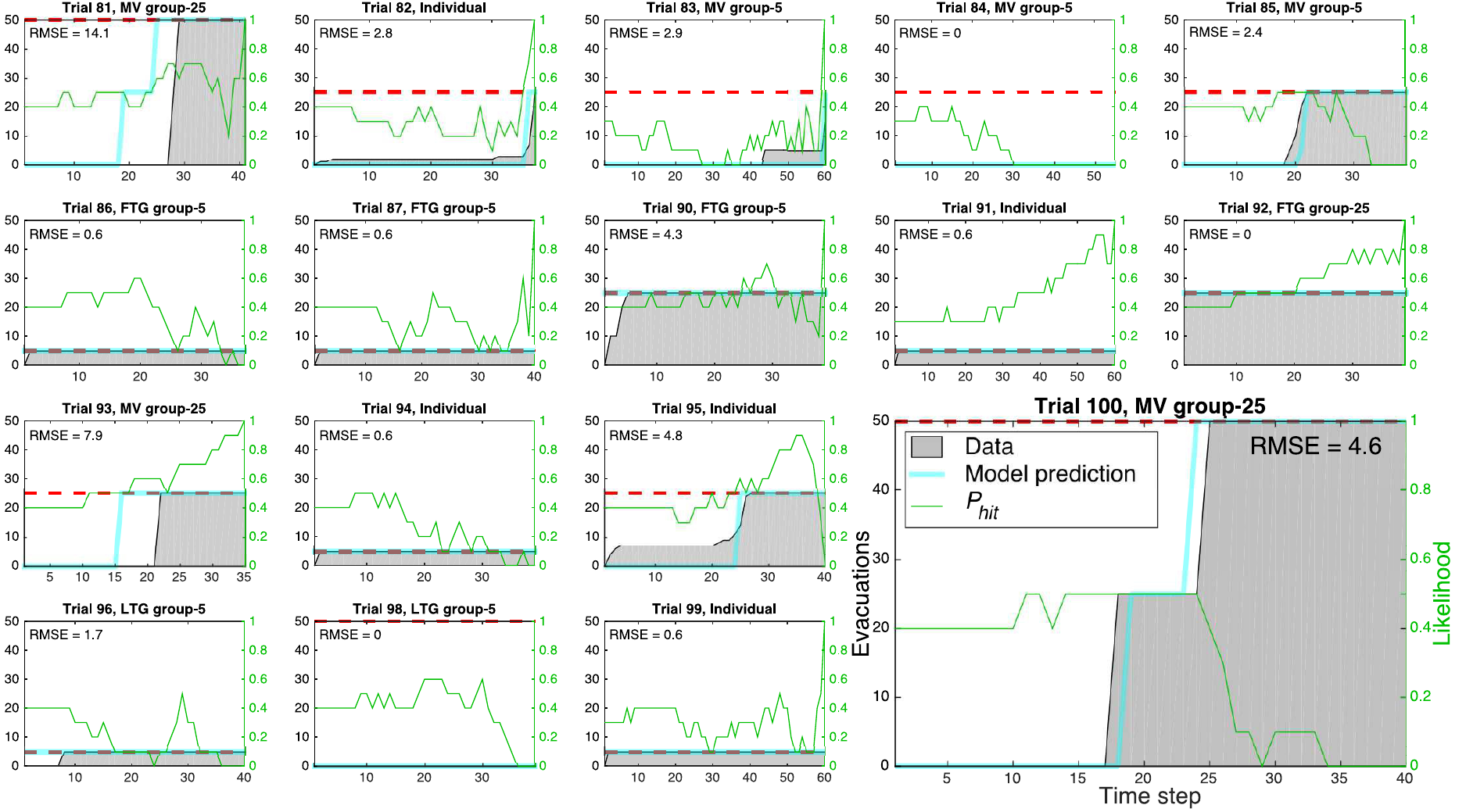}
\caption{\textbf{Comparison of observed behavior and model prediction for 17 trials.} Plotted are the observed cumulative number of evacuations as a function of experiment time (gray area) for 17 trials and the model prediction (blue) having trained the model on the remaining 111 trials. The $P_{hit}$ trajectory is also plotted in green and scaled between 0 and 1. The shelter space capacity is indicated by the dashed red line. The root-mean-square error (RMSE) is reported for each trial, calculated at each time step and averaged over all time steps in a trial.}
\label{fig:compare}
\end{figure*}

We assess the \textit{evacuation decision prediction accuracy} as the percentage of total decisions (positive and negative) correctly predicted for the test set consisting of the 17 withheld trials. The accuracy is computed as an average over all data points in the test set. Results of the final model applied to the complete set of 17 test trials, in sequence and without up- or down-sampling, are illustrated in Fig \ref{fig:compare}. {The model predictions are used to simulate the cumulative evacuations as a function of experiment time, which are then compared with the observed evacuations.}

The overall prediction accuracy of the baseline model applied to the test set is 75.1\% and is illustrated in the first column marked BM in Fig \ref{fig:measure}. This forms the starting point for identifying features that improve the accuracy of our results. 

\textbf{Evacuation decision time delay.} In some cases, evacuation decisions are recorded at anomalously low values of $P_{hit}$. For example, Fig \ref{fig:phitvstime} contains data points suggesting that certain individuals decide to evacuate when $P_{hit}$ is as low as zero. A close inspection of the trials in which this behavior is observed indicates that $P_{hit}$ was significantly higher on the preceding time step, so that the recorded decision coincides with a sharp drop in likelihood. This suggests that accurate representation of the observations must incorporate time delay. 

While cognitively a participant may decide to evacuate at time step $t$ in a trial, there are situations in which the recorded decision time may be $t+1$. This delay may be the result of a reaction time delay associated with elapsed time between the participant's initial observation of $P_{hit}$ and other state variables and their click of the Evacuate button, or by a slight system delay in server-client communication. Since the state variables update on half-second time increments, the $P_{hit}$ value may increase or decrease during this period, leading to a positive data point (i.e., evacuation decision) for values of $P_{hit}$ that did not trigger the decision. 

To account for this, we apply a time delay correction (Fig \ref{fig:nnflow}A) to identify the most probable decision time step. A backwards shift by one time increment is applied to the data when a decision is recorded on a time step in which $P_{hit}$ has dropped by at least 0.2 compared to the preceding time step. 
Adding the time delay correction to the baseline model improves its accuracy to 78.0\%; this is illustrated in Fig \ref{fig:measure} in the column marked +TD.

\textbf{Gradients in likelihood and shelter space.} The next step (Fig \ref{fig:nnflow}A) in refining model accuracy involves incorporating additional input nodes to capture dynamic changes that occur during the trials. Unlike the baseline model for which the input parameter values reflect only the current system state (e.g., current $P_{hit}$ and available shelter space), these additional variables reflect how parameters are changing over time. In the experiment, increases in $P_{hit}$ and decreases in shelter space, especially when they occur rapidly, are observed to trigger evacuation decisions, even when the current value of $P_{hit}$ is below average for decisions, or the number of available shelter spaces is not restrictive.  

To describe likelihood gradients, we add three additional input nodes to the baseline model state vector $\textbf{x}_{(n,t)}$ to input the likelihood over a time window that includes three time steps immediately preceding the current time step.
The resulting input vector $\textbf{x}_{(n,t)}$ now contains an extra three dimensions representing the likelihood gradient, for a total of 12 components. For instance, at $t = 34$ in trial 19, the $P_{hit}$ values at the preceding time steps $t = 31, 32, 33$ are all $0.4$. The three new dimensions are $[P_{hit}(n,t) - P_{hit}(n,t-3),$ $P_{hit}(n,t) - P_{hit}(n,t-2),$ $P_{hit}(n,t) - P_{hit}(n,t-1)]$, such that $\textbf{x}_{(n=19,t=34)} =  [0.5,31,1,0,0,0,0,0,0,0.1,0.1,0.1]$.

To describe changes in the available shelter space, we incorporate the shelter capacity at one time step immediately preceding the current time step as input to the model. Furthermore, as illustrated in Fig \ref{fig:spaceTime}, behavior differs drastically with the initial shelter capacity. Hence, we also use the total shelter capacity as input to the model. 
Two additional dimensions are thus added to the input vector $\textbf{x}_{(n,t)}$. One represents the total shelter capacity, and the other is the shelter capacity at $t$ minus the capacity at $t-1$. For example, in trial 19, the total capacity is 50, the shelter capacity at $t=34$ is 31, and the shelter capacity at $t=33$ is 34. The 14-dimensional input vector at $t=34$ is now $\textbf{x}_{(n=19,t=34)} = [0.5,31,1,0,0,0,0,0,0,0.1,0.1,0.1,50,-3]$. Adding the likelihood and shelter space gradients to the baseline model with time delay corrections, we obtain an accuracy of 80.2\%, as illustrated in Fig \ref{fig:measure} in the column marked +G.

\textbf{Individual differences.} We observe significant heterogeneity in behavior across individuals, as illustrated in Fig \ref{fig:phitvstime}. Some participants tend to decide early, even when the disaster likelihood is low, while other participants have greater spread in their decision times. Incorporating individual differences is the final step in developing our model. We utilize two different approaches: one is based directly on the experimental data, and the other is extracted from the participant's Facebook archive. Ultimately, we find that these two methods lead to comparable results. 
The experiment-based method adds a personality node (Fig \ref{fig:nnflow}B) and trains the model on the balanced data to obtain a unique personality factor for each participant. Specifically, the personality input is incorporated as a 1-out-of-$N$ set of nodes, where $N=50$, itemizing the number of participants, and the personality factor is the weight $w_i$ for individual $i$ obtained by training the networks. The accuracy obtained by adding the personality node to the baseline model with time delays and gradients is illustrated in Fig \ref{fig:measure} in the column marked +P. We obtain an accuracy of 85.0\%, a substantial improvement over the 75.1\% accuracy of the baseline model. This method is sufficient to capture individual personality differences and ultimately provides the most accurate prediction of collective behavior without overfitting.

As a separate approach to capturing individual heterogeneity, we aggregated each participant's Facebook archive information (age, gender, friends, and likes) to generate individual personality factors (Fig \ref{fig:nnflow}C) by adding an extra layer to the neural network model. However, strong L2 regularization was required to prevent overfitting, even when we restricted our use of Facebook likes to the most popular genres in training the model. An L2 regularization term was added to all weights corresponding to the Facebook features, and the strength of the regularization was determined by validation. This method ultimately provides similar accuracy to that obtained using the personality node determined directly from the data. We obtain an accuracy of 84.2\%, compared to 85.0\% obtained from the experiment-based personality model, as illustrated in in Fig \ref{fig:measure} in the column marked +FB.

For the model with personality parameters included, the decision probability $\mathcal{P}_{dec}(p,n,t)$ is unique for participant $p$ in trial $n$ at time $t$. The input vector $\textbf{x}_{(p,n,t)}$ is generated by concatenating all of the feature representations described above. The disaster likelihood and likelihood gradient, current and total shelter capacity, shelter gradient, group size, and group protocol, and their corresponding weights, are identical for every participant at a given time $t$ in a trial $n$, while the personality parameter is unique to each participant and is not time- or trial-dependent. 
Ultimately, the probability of participant $p$ deciding to evacuate at time $t$ in trial $n$ is defined by
\begin{equation}
\mathcal{P}_{dec}(p,n,t)=f(\textbf{w} \cdot \textbf{x}_{(p,n,t)}+b),
\label{eq:personalitymodel}
\end{equation}
where \textbf{w} represents the weighting of each dimension of the vector, and $b$ is the bias term. The function $f$ is the sigmoid function defined in Eq \eqref{eq:sigmoid}. 

\begin{figure*}[ht] 
\centering 
\includegraphics[scale=1]{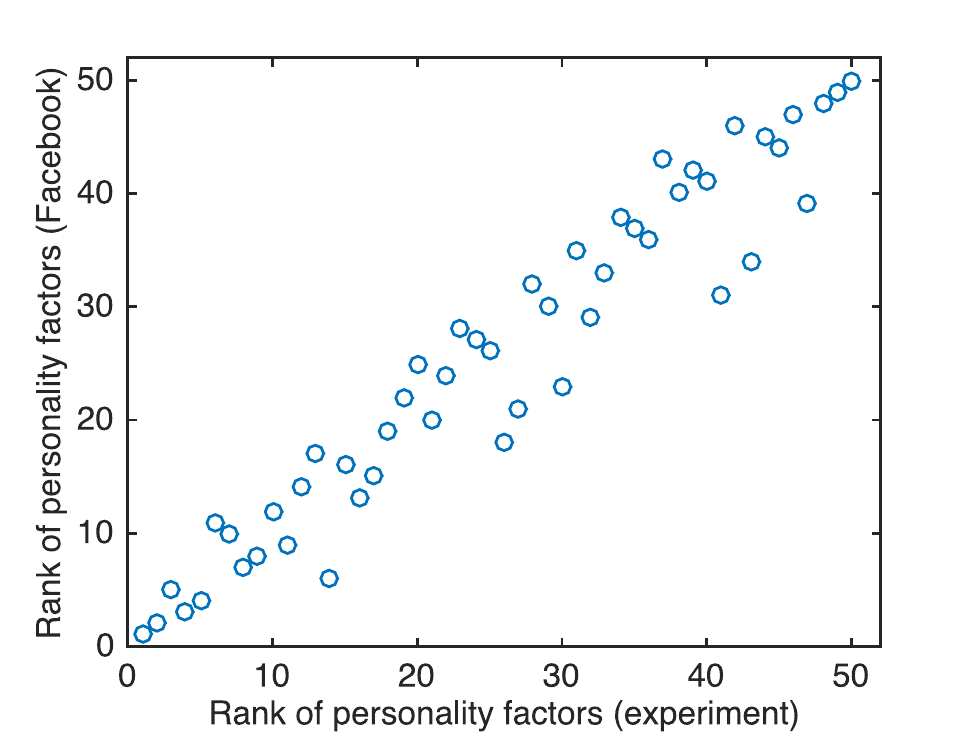}
\caption{\textbf{Comparison of personality factors.} The personality factors determined from training the model on experimental data (x-axis) are compared with those determined from Facebook data (y-axis). We plot the rank of the personality factors rather than the actual numerical values themselves, since the bias term $b$ will remove any offsetting. The Spearman rank correlation coefficient is 0.96 with $p < 0.001$.}
\label{fig:personality}
\end{figure*}

Additionally, as illustrated in Fig \ref{fig:personality}, we directly compare personality factors determined from training on the experimental data with those determined from Facebook data and find a strong correlation (Spearman rank correlation coefficient = 0.96, $p < 0.001$), indicating that Facebook demographic information and liked pages can be a good predictor of individual personality; however, note that the strength of the correlation may result in part from the strong regularization used when incorporating Facebook data into the model.

{Ultimately, with each incremental addition of features, the accuracy of the enhanced model increases relative to the baseline model, indicating that the additional features result in better prediction of the test set. Decisions to evacuate are dependent on current-time parameters as well as their immediate history and those that govern the situation on a more general level (total shelter capacity, group constraints). Since the test trials are withheld from the training data, and the test accuracy increases, the addition of features does not lead to overfitting of the training data.} 

\textbf{Comparing the model and experiment.} The final results of the artificial neural network model are illustrated in Fig \ref{fig:compare}, which compares the 17 test trials with results of the most accurate model (+P). The +P model combines the baseline model with the time delay correction, nodes accounting for gradients in likelihood and shelter space, and the individual personality node based on experimental data. Similar results are obtained for the +FB model, which replaces the experimentally determined personality node with information extracted from the participants' Facebook archives. 

The 17 test trials were selected from the middle portion of the experiment, and lie consecutively between trials 81 and 100, with three trials (88, 89, 97) omitted due to technical difficulties in recording data. We chose to use test trials near the middle of the experiment to minimize possible bias associated with early or late stages of the experiment. No particular attention was given to the type of trial; the sample includes at least one of every trial type except for FTG group-25 and LTG group-25.  
Model parameters were determined from the remaining 111 trials, with 80\% of the data (chosen randomly) used to fit parameters $\textbf{w}$ and $b$, and the remaining 20\% to optimize the threshold used to convert the continuous decision probability function $f(z)$ in Eq \eqref{eq:sigmoid} to a binary evacuate-or-not decision for each individual at each time step. 

Individual decision probabilities, as described in Eq \eqref{eq:personalitymodel}, are computed at each time step for the test data and converted to binary decisions for each participant. The decisions are then combined according to the assigned groups and protocols of the participants from the experiment to produce results for the model-predicted evacuations that are illustrated as blue curves in Fig \ref{fig:compare}. These predictions are made sequentially at each time step of the full data set containing 17 test trials, i.e., with no up- or down-sampling; the blue curves here depict only predicted group or individual evacuations, without showing the predicted decisions made by each group member. 
We observe striking agreement between the model predictions and the experimental data, represented as gray shaded regions. In most cases, the model accurately predicts not only the total number of evacuations, but also the approximate timing of the evacuations. The root-mean-squared error (RMSE) between observed and predicted evacuations is reported for each trial; the error is calculated at each time step and averaged over all time steps in a trial.

The trials exhibiting the greatest discrepancy between the model and observations are trials 81 and 93, both of which are MV group-25 trials. In both cases, the final total number of evacuations is correctly predicted; however, the model predicts that the evacuations will occur somewhat earlier than is observed (9 time steps earlier and 6 time steps earlier, respectively). 
Additional discrepancies are observed at early stages in some individual trials, such as trial 82 and trial 95, where a small number of participants evacuate in the first few time steps of the trial, even though $P_{hit}$ remains stationary and there are 25 shelter spaces. This early evacuation behavior may reflect the outcome of the previous trial, such as a disaster hit or shelter reaching capacity, which occurs in trials 81 and 94, respectively. Memory effects extending from one trial to the next are not captured by the model, which is based on single trial parameters.

\section*{Discussion}
We report results of a behavioral experiment investigating human decision making in the face of an impending natural disaster. We characterize individual and group behavior by quantifying several key factors for decision making in a controlled virtual community, including the disaster likelihood, shelter capacity, and decisions of group members, and we formulate a neural network model that predicts decision making of both individuals and groups in different scenarios. Including measures of individual personality significantly increases prediction accuracy compared to a baseline model dependent only on collective experimental variables. 

We find that individual behavior does not readily predict group behavior. We observe a significant difference in decision making between scenarios in which individuals decide solely for themselves and in which individual decisions are combined to determine collective action of a group. On average, participants decide to evacuate at lower values of the disaster likelihood when in groups than when acting as individuals. 

In many cases, we find that larger groups are less effective than smaller groups; the average success rate for individual trials is comparable to the average success rate of group trials, and groups of five perform better than groups of 25 under all protocols except majority-vote. The notion of the ``wisdom of crowds'' contends that the aggregated consensus decision of a collection of individuals is on average more accurate than an individual's decision \cite{munger} \cite{decisionsurvey} \cite{groupsize}. Condorcet's jury theorem states that, under the assumption that all individuals independently decide with an identical probability of correct decision greater than 0.5, a collective decision following the majority-vote protocol will have a greater success rate than an individual decision \cite{decisionsurvey}. In this idealized setting, the success of the majority-vote increases with group size and converges to 1 in the limit of infinite group size. An application of Condorcet's theorem states that if the individual success rate is less than 0.5, a pooled majority-vote decision will be less accurate than an individual decision, whereas if individual accuracy is greater than 0.5, the collective majority-vote accuracy increases as a function of individual accuracy and furthermore increases more sharply for larger group sizes \cite{groupaccuracy}. While our experiment is significantly more complex than this idealized voting scenario, on average, our observations from majority-vote trials remain consistent with Condorcet's theorem; the individual success rate is 0.604 (Fig \ref{fig:prinProb}), the group-5 MV success rate is 0.607, and the group-25 MV success rate is much higher at 0.75.

Including additional social factors such as the rankings of group members can give rise to a wide range of outcomes in group decision making \cite{munger}. If some individuals have a higher success rate than others, assigning greater weight to the most successful individuals' decisions when aggregating opinions into a consensus decision can result in greater collective accuracy \cite{decisionsurvey}. Furthermore, in more complex situations where individuals must decide in response to multiple cues with variable reliability, the wisdom of crowds effect is not necessarily observed; rather, there can be a finite optimal group size \cite{groupsize}. Ultimately, we do not find that larger groups necessarily perform better than smaller groups, and in many cases, groups of 25 are on average less successful than individuals.

While there is no significant overall correlation between a participant's rank in individual trials and group trials, we observe a significant correlation between individual rank and the time order of decisions in the LTG and MV group-25 trials that differentiates the top-performing group of 25 from the bottom-performing group. In the top-performing group, the higher-scoring participants decide earlier on average than the lower-scoring participants within the group, suggesting possible emergent ``leader'' and ``follower'' behavior. This correlation is not observed in the bottom-performing group. 

Given knowledge of the current state of the underlying stochastic process driving the disaster likelihood, it is possible to determine the optimal decision strategy \cite{phit}. In many scenarios, we observe behavior that is clearly sub-optimal. For example, when there is sufficient shelter space for all participants and the trial has just begun, some participants nevertheless decide to evacuate in the first few time steps, despite the lack of urgency in space or time. In those situations, participants would not experience a tradeoff between the information gained with time and the urgency to evacuate due to resource competition. In some cases, we also observe extreme trial-to-trial variations, such as early evacuations in the absence of time or space constraints, following trials in which the shelter reaches capacity or the disaster hit. Our results emphasize the importance of properly characterizing sub-optimal decisions and non-Bayesian adjustments in strategy over time in developing predictive models of human decision-making behavior. The sub-optimality of real human decision making is explored in \cite{paper2}, which compares observed behavior with the optimal strategy as determined under the assumption of a perfect ``Bayesian learner'' who aggregates accumulated evidence from previous experience to inform decisions. 

The modeling procedure performed in this paper enables the identification and quantification of key factors, using machine learning techniques to identify correlation structure, while also taking into account individual differences in behavior which can be extracted from social network data. This work also motivates the development of a model for individual decision making in \cite{paper2} that is based on a few key parameters and contains an explicit functional form derived from the statistics of experimental data. That model accurately predicts individual decision making and is used to simulate group behavior, forming a baseline of predictions that are contrasted with observed group behavior to examine the differences between individuals and groups.

\textbf{Relationships to past work.} Our work builds on four prior studies of collective behavior in evacuation scenarios. The first is a spatiotemporal analysis of wildfire progression on a rural landscape and transportation network that determines optimal evacuation routes and clearing times in a realistic setting under the assumption that all individuals act optimally \cite{langford}. While the decision model is idealized and hence unrealistic, it provides an upper bound on the success of collective behavior. 

The second study developed a stochastic model to simulate the progression of a threat by characterizing the evolution of the strike probability as a Markov process, and determined optimal decision strategies in the face of stochastically-varying uncertainty \cite{phit}. We applied this stochastic process to determine the progression of disaster likelihood for each disaster scenario in our experiment. 

The third study simulated the decision dynamics of individuals on a social network, where decisions were influenced both by uniform global information broadcast and pairwise neighbor interactions \cite{dani}. This study found that information transmission on a network could either facilitate or hinder collective action, depending on the influence of social network interactions, and identified the mechanisms leading to cascading behavior and stagnation. 

The experimental platform in this paper is largely derived from the fourth study, in which a similar controlled behavioral experiment involving an evacuation scenario was performed, and from which a data-driven model was developed to characterize the influence of disaster likelihood, time pressure, and shelter capacity on decision making \cite{sean}. Compared to the previous study, here we simplify the presentation of broadcast information (time progression and disaster likelihood) and social information (decisions of others and shelter space) so that it appears on a single screen, rather than being split into separate broadcast and social tabs which participants could switch between during each trial. The previous study also included the state of being ``in transit'' in addition to home or shelter; if a participant decided to evacuate when the shelter was full, they would remain in the ``in transit'' state for the remainder of the trial. The losses incurred from being in the ``in transit'' state were an intermediate value between those of home and shelter. In the previous experiment, the loss matrix varied from trial to trial and was displayed on the computer interface. However, values in the loss matrix were found to have minimal impact on decision making in the earlier study. Therefore, for the current study we chose to fix the loss matrix and only displayed it briefly during the introductory training session. This allowed us greater flexibility to vary other parameters in the nested experimental design.

Our experiment differs most significantly from this previous study with the introduction of forced group trials and protocols for group action. The previous experiment allowed participants to probe the decisions of their neighbors over a structured social network. In this manner, individuals could gain information about the decisions of others and the remaining shelter space; however, each participant acted alone in every trial. This allowed for the possibility of clustered evacuations, but did not force groups to act together. 
In the earlier study, participants spent most of their time on the broadcast information tab and relatively little time on the social information tab, aside from the start of the experiment where the available shelter space was displayed. This motivated our design of a group evacuation experiment to target the role of social information as a driver of decision making. 

Individual personality, preferences, and risk perception have been shown to influence decision making in evacuation scenarios \cite{evachouse}. In the current study, we quantified individual personality factors by training an artificial neural network containing a ``personality node'' on the data and correlated our model with personality factors determined from Facebook activity. In contrast, the previous study employed written surveys to determine personality factors, specifically the Big Five Inventory questionnaire, derived from the five-factor (extraversion, agreeableness, conscientiousness, neuroticism, and openness) model \cite{bigfivedigman} \cite{bigfivemccrae}, and the Domain-Specific Risk-Attitude Scale, a risk perception survey specific to six domains (social, investment, gambling, health and safety, ethical, and recreational) \cite{risk1}\cite{risk2}. 

The popularity of online social networks such as Facebook and Twitter has motivated the investigation of social network data to uncover predictive measures of individual traits \cite{fb} \cite{fblikes}. Social network data provides a potentially more direct assessment of ongoing activity compared to self-reported preferences. Furthermore, there is significantly more information in social network data (which is often unavailable due to proprietary or privacy reasons) which, when properly filtered, may further refine predictions of group relationships and individual personality differences. 

\textbf{Methodological considerations.} {We made multiple intentional choices in the presentation of broadcast information viewed by the participants. For instance, we displayed $P_{hit}$ as a coarse-grained likelihood rather than a probability, since the presentation of probabilistic information has been shown to affect resultant decisions \cite{edwards}. We also represented risk information as losses rather than incentives to reflect the focus on loss of life or property rather than gains in real disaster situations. Studies have shown that humans react differently to losses than to gains, assigning greater value to differences of equal magnitude in losses compared to gains \cite{prospect} \cite{prospect2}. Moreover, when performing incentivized tasks, individuals encode the potential loss due to failure rather than the potential gain arising from success, causing some individuals' performance levels to decline as incentive levels increase \cite{chib}.} 

The behavioral distinction in loss-based versus gain-based experimental settings can be extended to a distinction between risk-based and profit-based scenarios. Risk scenarios such as the natural disaster scenarios considered in this experiment also include other situations that directly endanger personal safety, such as battlefield or military scenarios, firefighting, and search and rescue missions. On the other hand, profit scenarios include group decision making in the workplace, community organizations and committees, or foraging (in animal communities).

We have chosen to ignore spatial variation in information arrival times, focusing instead on uniformly distributed information regarding a disaster that impacts or misses the entire community at once. This isolates key tensions between individual and group decision making in an otherwise homogeneous context. In the case of spatial search or monitoring, which is also observed in collective animal behavior (e.g., fish schooling), increasing group size is beneficial as the group can cover more area, unless the area or another resource necessary to perform the action is limited. For instance, the collective navigation and environmental sensing performance of golden shiners increases monotonically with group size, since individual fish rely more strongly on observations of neighboring fish rather than on the environment itself \cite{fishgroups}.

We also fixed the decision protocols (FTG, LTG, and MV) in order to isolate the effect of different protocols on behavior. A popular choice of decision protocol is majority-vote, which has been observed in both hunter-gatherer tribal societies and modern democracies \cite{majority}. In laboratory settings, when groups are free to act according to any decision rule, groups tend to choose the majority-vote protocol if information broadcast is uniform across all group members \cite{eckstein}. However, if information broadcast is not uniform, groups adjust their decision protocol, often following a ``majority with exceptions'' rule, where majority-vote is followed unless one or more group members has an opposing opinion with a high confidence level.

In our experiment, participants are assigned to groups without their input. Other types of ``forced'' groups that have been defined in advance include workplace teams and military units. However, individuals may still refuse to comply with the rest of the group or ``opt out'' of group action, e.g., going AWOL from a military base. Future experiments will explore the contrast between ``forced'' group behavior with the naturally-arising collective behavior of ``self-assembling'' groups of humans or animals (e.g., fish schools and bird flocks). 

The participants of this experiment were mostly undergraduate students, the majority of whom were male (Fig \ref{fig:facestats}). Evacuation decisions are highly influenced by age, socioeconomic status, geographic location, household size, and responsibility for children (the presence of which tend to encourage evacuation) or the elderly (which discourage evacuation) \cite{evachouse} \cite{gladwinpeacock}. Risk perception also varies with age, cultural background, and experience \cite{evachouse}. Future studies will target participants from different demographic backgrounds in order to further probe the relationship between demographics and decision making.

\textbf{Concluding remarks.} In natural and man-made disasters, information on the progression and magnitude of an impending disaster is distributed through a combination of global broadcast networks (e.g., television, radio, and the Internet) and social networks (e.g., Facebook, Twitter, and text messaging). The former is generally associated with news organizations that present information in a formal, edited manner, while the latter is generally based upon informal direct observations made by individuals in the population. 

Increasingly, the speed of information flow over social networks outpaces traditional broadcast communications, creating new opportunities to inform the public at unprecedented rates. However, along with fast-paced updates, social media viewed as a disaster communication tool introduces new and potentially dangerous fragilities associated with asynchronous, non-uniform, and incorrect updates that are intrinsic to unfiltered, open-access communication, yet are currently not well characterized. Future extensions of our experimental platform will address these issues by incorporating asynchronous information updates within the population, spread and detection of misinformation, self-assembly and disassembly of groups, and realistic topography and transportation routes in order to develop increasingly realistic scenarios for human decision modeling.

Although this paper details a relatively small-scale experiment in a controlled and simplified setting, our findings may be extended to guide the development of policies for collective action in more complex situations of stress and uncertainty.  Decisions to evacuate in the face of impending disaster may depend not only on the current weather forecast, for example, but also on the volatility of the weather conditions over the past several hours, as reflected by our model, which takes into account both current parameters and those within a recent time window, as well as other trial-dependent factors such as overall shelter capacity and group size. Moreover, the prevalence of sub-optimal decisions and non-Bayesian strategy adjustments over time observed in our experiments serves as a precautionary guide to the development of large-scale, agent-based simulations of evacuations. This paper also addresses the limitations of extending characteristics of individuals without consideration of social influence to predict the behavior of groups. Our results indicate significant differences in behavior between group sizes and protocols, and our observation that individual decision making does not generally predict strategy or rank in group action mandates the incorporation of social factors in computational modeling and design of optimized infrastructure and action protocols in preparation for natural or man-made disasters.

\nolinenumbers

\section*{Acknowledgements}
This work was supported by the David and Lucile Packard Foundation, the National Science Foundation Graduate Research Fellowship Program under grant no.\ DGE-1144085, and the Institute for Collaborative Biotechnologies through contract no.\ W911NF-09-D-0001.

\end{document}